%% file: 1850.tex
\documentclass{aa}
\usepackage{txfonts}
\usepackage{graphicx}

\begin{document}
    \title{A deep 6.7\,$\mu$m survey in the SSA13 field with ISO
           \thanks{Based on observations with ISO, an ESA project
             with instruments funded by ESA Member States (especially
             the PI countries: France, Germany, the Netherlands and
             the United Kingdom) and with the participation of ISAS and NASA.}
          }
    \author{Y.~Sato\inst{1,2}
       \and K.~Kawara\inst{2}
       \and L.~L.~Cowie\inst{3}
       \and Y.~Taniguchi\inst{4}
       \and D.~B.~Sanders\inst{3,7}
       \and H.~Matsuhara\inst{1}
       \and H.~Okuda\inst{5}
       \and K.~Wakamatsu\inst{6}
       \and Y.~Sofue\inst{2}
       \and R.~D.~Joseph\inst{3}
       \and T.~Matsumoto\inst{1}
           }
    \offprints{Y.~Sato,
          \email{ysato@ioa.s.u-tokyo.ac.jp}
          }
    \institute{Institute of Space and Astronautical Science (ISAS),
                 3-1-1 Yoshinodai, Sagamihara, Kanagawa, 229-8510 Japan
          \and Institute of Astronomy, University of Tokyo,
                 2-21-1 Osawa, Mitaka, Tokyo, 181-0015 Japan
          \and Institute for Astronomy, University of Hawaii,
                 2680 Woodlawn Drive, Honolulu, HI 96822, USA
          \and Astronomical Institute, Graduate School of Science, 
Tohoku University,
                 Aramaki, Aoba, Sendai, 980-8578 Japan
          \and Gunma Astronomical Observatory,
                 6860-86 Nakayama, Takayama, Agatsuma, Gunma, 377-0702 Japan
          \and Faculty of Engineering, Gifu University,
                 1-1 Yanagido, Gifu, 501-1193 Japan
          \and Max-Planck Institut fur Extraterrestrische Physik,
                 D-85740, Garching, Germany
              }
    \date{Received ; accepted }

\abstract{
We present results of a deep mid-infrared survey in the SSA13 field
with the Infrared Space Observatory (ISO).
In order to probe the near-infrared light at high redshifts,
we surveyed the field with the broad band LW2 (5--8.5\,$\mu$m) filter
of the mid-infrared camera ISOCAM.
Adopting a highly redundant imaging strategy for the 23 hour observation
and carefully treating gradual changes in the detector responsivity
caused by a very high rate of cosmic ray impacts,
we succeeded in reaching an 80\,\% completeness limit of 16\,$\mu$Jy
in the central 7 arcmin$^2$ region.
Utilizing the signal-to-noise ratio map,
we detected 65 sources down to 6\,$\mu$Jy in the 16 arcmin$^2$ field.
Integral galaxy number counts at 6.7\,$\mu$m are then derived,
reaching $1.3\,\times\,10^4$ deg$^{-2}$ at the faint limit with a slope of
$-1.6$
between 13\,$\mu$Jy and 130\,$\mu$Jy.
Integrating individual sources in this flux range,
the resolved fraction of the extragalactic background light at 6.7\,$\mu$m
is estimated to be 
0.56\,nW\,m$^{-2}$\,sr$^{-1}$.
These results, which reach a flux limit three times fainter than those in the
Hubble Deep Fields, are in fairly good agreement with a model prediction
by Franceschini et al.\ (1997).  Finally, we discuss the relation of 
distant massive
E/S0 galaxies to the faint 6.7\,$\mu$m galaxy population.
}
\maketitle


\section{Introduction}

Explorations of the evolution of galaxies
require representative galaxy samples at high redshifts.
Intrinsically sensitive optical observations
succeeded in picking up $z \sim 3$ sources
utilizing characteristic breaks in galaxy spectra
(Steidel et~al. \cite{SGP+96}).
Another technique to search for high redshift galaxies
is to observe at wavelengths long-ward of a global peak
in spectral energy distributions (SEDs) of galaxies.
A famous example is the recent discovery
of distant dusty starbursting galaxies
in the Submillimeter and millimeter
with the Submillimeter Common-User Bolometer Array SCUBA
(Smail et~al. \cite{SIB97};
Barger et~al. \cite{BCS+98};
Hughes et~al. \cite{HSD+98})
and the Max-Planck Millimeter Bolometer array MAMBO
(Bertoldi et~al. \cite{BCM+00}).
At submillimeter and millimeter wavelengths,
contributions from foreground objects are small, since
their dust emission has a peak
around 100\,$\mu$m and decreases toward longer wavelengths.
On the other hand,
contributions from higher redshift objects become greater
because of the K-correction brightening;
the rest-frame 100\,$\mu$m dust peak shifts into
the submillimeter or millimeter observing band
with increasing redshift.
As a result,
the fraction of high redshift objects in the submillimeter or millimeter
is greater than that at, say, 200\,$\mu$m.

The same technique can be applied to
another striking peak around 1\,$\mu$m originating from stellar photospheres.
The observing wavelength should be much longer than 1\,$\mu$m
to get the K-correction brightening.
However,
emission from PAH molecules and very hot dust should also be considered.
Their contributions begin to increase at wavelengths long-ward of 6\,$\mu$m
for local objects.
We selected the broad band LW2 filter at 6.7\,$\mu$m in the 
mid-infrared camera ISOCAM
(Cesarsky et~al. \cite{CAA+96})
as a good window between local stellar and dust emission.
The fraction of high redshift objects
would be larger at 6.7\,$\mu$m than at 2.2\,$\mu$m.

Previous field galaxy surveys at 6.7\,$\mu$m
were carried out
in the Hubble Deep Field (HDF) and the Lockman Hole
(Serjeant et~al. \cite{SEO+97};
Taniguchi et~al. \cite{TCS+97}).
These found a few tens of 6.7\,$\mu$m sources down to 30\,$\mu$Jy
(Goldschmidt et~al. \cite{GOS+97}), some of which were confirmed by
re-calibration of the HDF data
(D{\'e}sert et~al. \cite{DPC+99};
Aussel et~al. \cite{ACE+99}).
A more shallow ($>100\,\mu$Jy) survey was conducted in the CFRS field
(Flores et~al. \cite{FHD+99}).
An attempt to detect a field population lensed by a massive cluster lens
was also performed at 6.7\,$\mu$m
(Altieri et~al. \cite{AMK+99}),
reaching a faint detection limit comparable to
those of the HDF and the Lockman Hole.
Most recently, a survey of the HDF-South
(HDF-S; Oliver et~al.\ \cite{OMC+02})
has been reported with the same depth as that of the HDF.

Here we present a very deep ISOCAM LW2 galaxy survey in the SSA13 field.
We first describe the observing strategy (Sect.~\ref{sect:obs})
and the data processing (Sect.~\ref{sect:image}).
After evaluating the detection procedure by simulations
(Sect.~\ref{sect:obj_ex}),
we present the catalog of detected sources (Sect.~\ref{sect:cat}).
Finally we discuss
the galaxy number counts and extragalactic background light
(Sect.~\ref{sect:count}).
Other properties of the individual sources will be described
elsewhere (Sato et~al. \cite{SCK+03}).


\section{Observation}
\label{sect:obs}

\subsection{Field selection}

Small Selected Area 13 (SSA13), one of the Hawaii Deep Fields,
was chosen as our deep 6.7\,$\mu$m survey target.
It is located at high galactic and ecliptic latitudes
($b \sim 74\degr$ and $\beta \sim 46\degr$)
and has a very low atomic hydrogen column density
of $\mathrm{N(\ion{H}{i})} = 1.5\,\times\,10^{20}$ cm$^{-2}$
(Cowie et~al. \cite{CGH+94}).
Owing to these attractive properties for extragalactic studies,
SSA13 has become one of the most-studied regions of the sky,
comparably with the HDFs.
It has photometric data obtained not only in the optical and near-infrared,
but also at submillimeter and X-ray wavelengths
(Cowie et~al. \cite{CSH+96};
Barger et~al. \cite{BCS+98}, \cite{BCS99};
Mushotzky et~al. \cite{MCB+00}).
A number of spectra were taken
with the Low Resolution Imaging Spectrometer (LRIS) of the Keck telescope
in a certain portion of the SSA13 field
(Cowie et~al. \cite{CSH+96}).
Because of the presence of a bright star ($R=10.4$),
the survey area was chosen to be the largest square region
which is most distant from the star but inside the rectangular LRIS strip.
The center of the region is (RA, Dec)$=$
($13^\mathrm{h}\,12^\mathrm{m}\,$26\fs0,
42\degr\,44\arcmin\,24\farcs8) in J2000.

\subsection{Observing strategy}
\label{sect:strategy}

The survey observations were conducted with the mid-infrared camera ISOCAM
(Cesarsky et~al. \cite{CAA+96})
on board the Infrared Space Observatory ISO
(Kessler et~al. \cite{KSA+96}).
The sensitive broad band LW2 (5--8.5\,$\mu$m) filter
with a reference wavelength 6.7\,$\mu$m
(Moneti et~al. \cite{MMS97}) was selected
to detect radiation from stellar photospheres at high redshift.
All the images were taken in the microscanning mode, CAM01,
to minimize the flatfielding error
(Siebenmorgen, Sauvage \& Levine \cite{SSL96}).
Total target-dedicated-time (TDT) for the observations was 23 hours.

Because of the visibility constraints for the SSA13 field,
our observations were divided over six revolutions
between revolution 560 and 614,
29 May 1997 and 22 Jul 1997 in UT
(Table~\ref{tab:rev}).
The observing time in each revolution
was divided into two or three raster observations
(Table~\ref{tab:raster}).
The observations were executed
essentially in two runs separated by a 1.5 month interval,
during which time the image plane of the spacecraft rotated about 45 degrees
({\it cf.} roll angles in Table~\ref{tab:rev}).
The main aim was to achieve better sampling of source profiles
and also to maximize the overlapped area coverage.

The surface density of bright sources suitable for image registration
was expected to be low at 6.7\,$\mu$m.
Thus, it was almost mandatory to fully utilize
the unvignetted field-of-view for ISOCAM, 3 arcmin in diameter.
We then selected the lens providing a 6 arcsec pixel field-of-view
on the $32 \times 32$ pixel LW detector array.
However,
the use of 6 arcsec lens could introduce errors in the registration
and hence degrade the depth of the map, because 6 arcsec pixels undersample
the instrumental beam.
At 6.7\,$\mu$m, the diffraction limited beam size is 2.3 arcsec
FWHM (Full Width at Half Maximum)
for the 60\,cm ISO mirror.
In order to suppress this undersampling effect,
we applied several raster step sizes
corresponding to non-integer multiples of the detector pixel size
(Table~\ref{tab:raster}).

Because of the likely faintness of sources that could be used
as references for the image registration,
all the raster observations in a single revolution were executed
as a concatenated chain.
If we had not adopted this concatenation setting,
we could not align the CAM images and it would result in null results.
In the concatenated chain, the nominal automatic transition
to the standby mode between observations was suppressed.
That transition, if permitted, would have introduced
  strong transients in the detector
responsivity and random and systematic displacements of the image on 
the detector
(Siebenmorgen et~al. \cite{SSL96}).
In fact, non-negligible displacements were actually identified
among revolution maps taken with different lens positions
(Table~\ref{tab:rev}).

The integration time per exposure was set to T$_{\mathrm{int}}=20$\,sec.
With shorter integration time,
data would have been dominated by the readout noise of the detector
because of very low background emission at 6.7\,$\mu$m.
The number of exposures per raster position was chosen to be 
N$_{\mathrm{exp}}=12$,
following a caveat for robust cosmic ray rejection
(Siebenmorgen et~al. \cite{SSL96}).
The number of stabilization exposures was set to N$_{\mathrm{stab}}=21$
for the first raster in each revolution.
The stabilization exposures were used as a buffer
for the associated strong transients from the standby mode.
Here the worst case was assumed,
in which an observation would be preceded by a dark (zero flux) condition.
For the subsequent rasters in any given revolution,
the minimum stabilization value N$_{\mathrm{stab}}=4$ was used, since
no change in
background illumination would occur in the concatenated chains. The 
analog-to-digital
converter gain was set to the maximum $\times 4$ to cope with the low 
contrast of faint
sources on the background.

\begin{table*}
   \centering
   \caption[]{
   Properties of the ISOCAM observations in the individual revolutions.
     Column 1 is the revolution number counted since ISO's launch.
     Column 2 is the total target-dedicated-time (TDT) in a single revolution.
     Column 3 is the roll angle of the satellite,
              counted counterclockwise from north to the raster N-direction.
     Column 4 is the solar aspect angle between the target and the Sun.
     Column 5 is the measured sky brightness derived with the 
sensitivity parameter of
              $2.32 \pm 0.08$\,ADU/G/s/mJy for the LW2 band (Blommaert 
\cite{B98}).
              ISOCAM signals in units of ADU/G/s (Analog-to-Digital 
Unit / Gain / second)
              can be converted to flux densities with the sensitivity 
parameter of the
filter used.
     Column 6 is the lens position following the definition by
              Aussel et~al. (\cite{ACE+99}).
     Columns 7 and 8 are the measured offsets between the ISOCAM 
detector and the nominal
              center of the ISOCAM field-of-view determined by the telescope.
              They are shown in the raster M- and N-directions in 
units of pixels.
              These shifts were caused by the jitter in the lens wheel 
repositioning
              (Sect.~\ref{sect:strategy}).
     }
   \label{tab:rev}
   \catcode`?=\active \def?{\phantom{0}}
   \begin{tabular}{cccccccc}
       \hline
       \noalign{\smallskip}
Revolution & Observing time & Roll angle & Solar aspect angle & Sky 
& Lens position & Offset in M & Offset in N \\
            & (sec)          & (deg)      & (deg)              & 
(MJy\,sr$^{-1}$) &               & (pix)       & (pix)       \\
       \noalign{\smallskip}
       \hline
       \noalign{\smallskip}
560 & 11214 & 315.2 & 102.0 & 2.90 & left  & $?0.95$ & $-0.24$ \\
561 & 11214 & 314.4 & 101.4 & 2.94 & right & $-1.10$ & $-0.10$ \\
610 & 15315 & 279.9 & ?69.2 & 4.11 & left  & $?1.21$ & $?0.16$ \\
612 & 15315 & 278.4 & ?68.1 & 4.21 & left  & $?0.90$ & $-0.12$ \\
613 & 15315 & 277.6 & ?67.4 & 4.28 & right & $-0.97$ & $?0.19$ \\
614 & 15315 & 276.7 & ?66.8 & 4.19 & right & $-0.80$ & $?0.14$ \\
       \noalign{\smallskip}
       \hline
   \end{tabular}
\end{table*}

\begin{table}
   \centering
   \caption[]{
Raster parameters used for the individual ISOCAM observations.
Column 1 is the TDT number, the unique identification number for ISO 
observations.
Columns 2 and 3 are the numbers of raster points
in the raster M- and N-directions.
The raster directions were fixed in the spacecraft coordinate system.
Columns 4 and 5 are the raster step sizes.
Note the pixel field-of-view was set to 6 arcsec.
   }
   \label{tab:raster}
   \begin{tabular}{ccccc}
       \hline
       \noalign{\smallskip}
TDTNR & M & N & $\Delta$M & $\Delta$N \\
       &   &   & (arcsec)  & (arcsec)  \\
       \noalign{\smallskip}
       \hline
       \noalign{\smallskip}
56000101 & 6 & 6 & 7 & 7 \\
56000102 & 3 & 2 & 9 & 9 \\
\hline
56100303 & 6 & 6 & 7 & 7 \\
56100304 & 3 & 2 & 9 & 9 \\
\hline
61000505 & 6 & 6 & 7 & 7 \\
61000506 & 4 & 4 & 8 & 8 \\
61000507 & 3 & 2 & 9 & 9 \\
\hline
61200108 & 6 & 6 & 7 & 7 \\
61200109 & 4 & 4 & 8 & 8 \\
61200110 & 3 & 2 & 9 & 9 \\
\hline
61300211 & 6 & 6 & 7 & 7 \\
61300212 & 4 & 4 & 8 & 8 \\
61300213 & 3 & 2 & 9 & 9 \\
\hline
61400314 & 6 & 6 & 7 & 7 \\
61400315 & 4 & 4 & 8 & 8 \\
61400316 & 3 & 2 & 9 & 9 \\
       \noalign{\smallskip}
       \hline
   \end{tabular}
\end{table}


\section{Image processing}
\label{sect:image}

\begin{figure}
   \centering
   \resizebox{\hsize}{!}{\includegraphics[width=\textwidth]{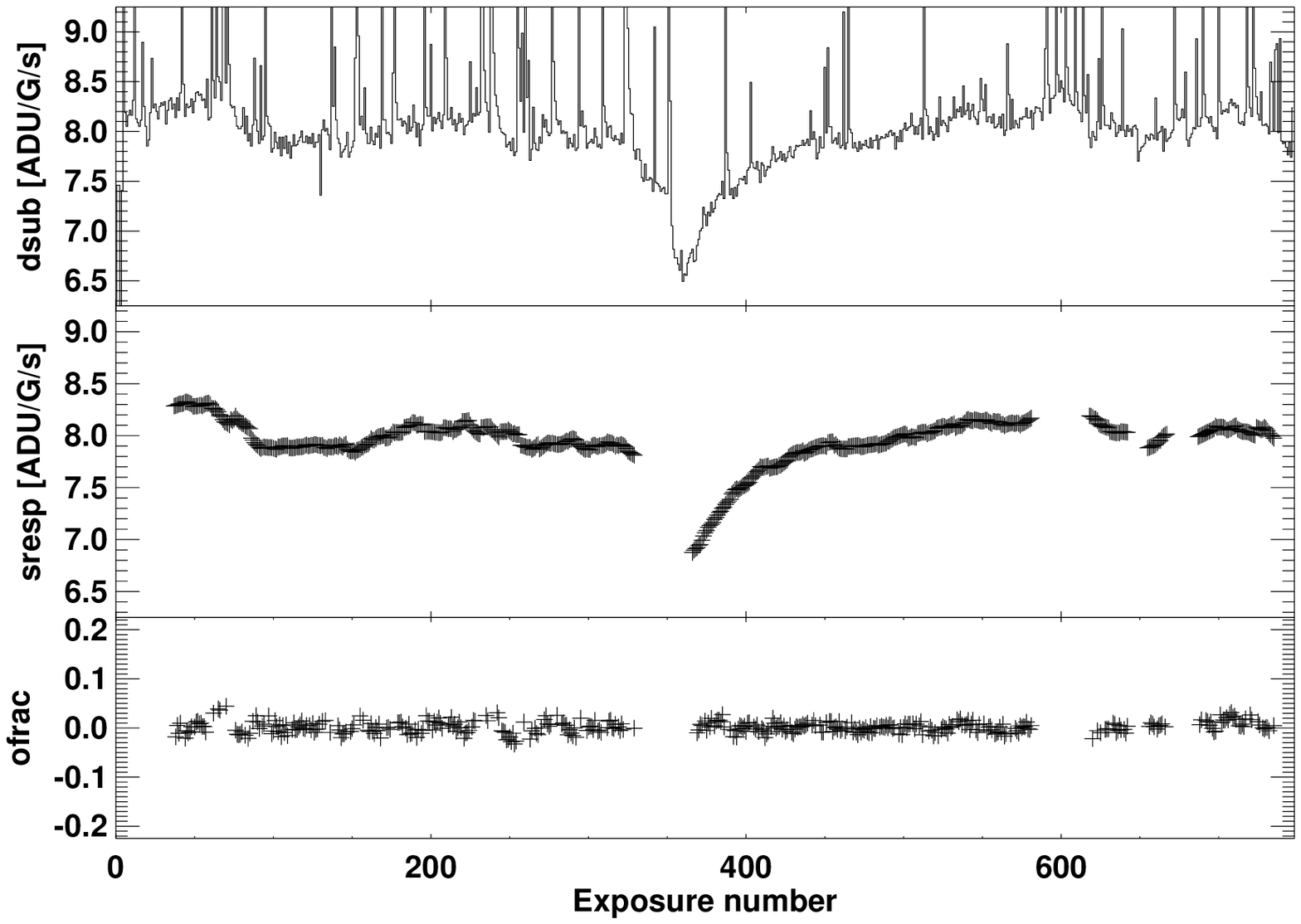}}
   \caption{
An example of the responsivity correction
to ISOCAM data taken with T$_{\mathrm{int}}=20$\,sec.
The top panel shows the dark-subtracted data
$dsub=a\,(I_{\mathrm{sky}}+I_{\mathrm{obj}})$.
The middle is $sresp=a\,I_{\mathrm{sky}}$,
the response to the blank sky after cosmic ray and object masking.
The bottom shows $ofrac=I_{\mathrm{obj}}/I_{\mathrm{sky}}$
after correcting the responsivity drifts.
For the $sresp$ and $ofrac$ panels,
bad portions are masked out.
The bad portions are cosmic rays themselves
and portions where good linear interpolations
are impossible with cosmic rays-masked and object-masked $dsub$ data
(Appendices~\ref{sect:rpmwom}, \ref{sect:crmask}).
   }
   \label{fig:history3}
\end{figure}

Mid-infrared data with ISOCAM suffer from cosmic rays
(top panel of Fig.~\ref{fig:history3}).
There is a huge number of spikes in the pixel histories
and the signals slowly change even when the background is constant.
The latter is attributed
to gradual changes or drifts in the detector responsivity
due to the large number of cosmic ray impacts.
Adopting the cosmic ray masking and the responsivity correction
to deal with the above effects,
we have done the data processing following the scheme outlined below.
Details of these steps are described in Appendix~\ref{sect:details}.
They are coded as our own software.

For each detector pixel,
its raw signal $raw$ can be written as the linear equation
\begin{equation}
	raw = a\, (I_{\mathrm{sky}} + I_{\mathrm{obj}}) + b	\label{eq:raw},
\end{equation}
where $a$ is the responsivity, $b$ is the dark signal,
and $I_{\mathrm{sky}}$ and $I_{\mathrm{obj}}$ are fluxes
from the sky and objects received by that pixel, respectively.
The dark-subtracted signal $dsub$
(top panel of Fig.~\ref{fig:history3}) is then obtained as
\begin{eqnarray}
	dsub	& = &	raw-b 
	\label{eq:dsub}	\\
		& = &	a\,(I_{\mathrm{sky}}+I_{\mathrm{obj}}).
\end{eqnarray}
The sky signal $a\,I_{\mathrm{sky}}$ can be extracted
by masking the object signal $a\,I_{\mathrm{obj}}$ and also cosmic rays
(Appendices~\ref{sect:objmask}, \ref{sect:crmask}).
This manipulation $f(dsub)$
is essentially a linear interpolation of $dsub$ signals
applying the object and cosmic ray masks.
The outcome reflects the detector response to the constant sky flux.
We call it $sresp$ (middle panel of Fig.~\ref{fig:history3}),
\begin{eqnarray}
	sresp	& = &	f(dsub)		\label{eq:a}		\\
		& = &	a\,I_{\mathrm{sky}},
\end{eqnarray}
and this will be used to correct the responsivity drifts.
Then, the time-dependent responsivity $a$
can be canceled out with the division of $dsub$ by $sresp$,
\begin{eqnarray}
	dsub/sresp	& = &	1+I_{\mathrm{obj}}/I_{\mathrm{sky}}.
\end{eqnarray}
Thus, the fraction of the object flux $ofrac$
can be given as
\begin{eqnarray}
	ofrac	& = &	I_{\mathrm{obj}}/I_{\mathrm{sky}}	\\
		& = &	dsub/sresp-1.	\label{eq:ofrac}
\end{eqnarray}
This technique for the responsivity correction works very well,
resulting in a drastic reduction of noise
({\it cf.} bottom panel of Fig.~\ref{fig:history3}).

Although $I_{\mathrm{sky}}$ is constant in time,
each detector pixel has a different $I_{\mathrm{sky}}$
due to the optical distortion
(Aussel \cite{A98})
and vignetting in the LW array.
Let the pixel
receive
$\alpha$ times larger sky flux than the fiducial pixel.
Then,
\begin{equation}
	I_{\mathrm{sky}} = \alpha\,I^0_{\mathrm{sky}}	\label{eq:alpha},
\end{equation}
where $I^0_{\mathrm{sky}}$ is the sky flux for the fiducial pixel.
The fiducial sky flux $I^0_{\mathrm{sky}}$
is derived from the central part of the detector
(Appendix~\ref{sect:sky}).
Then the $ofrac$ data are transferred to
the object flux $obj$ normalized to the fiducial pixel,
\begin{eqnarray}
	obj	& = &	ofrac \times I^0_{\mathrm{sky}} 
	\label{eq:obj}	\\
		& = &	I_{\mathrm{obj}}/\alpha.
\end{eqnarray}
Deriving the conversion factor for the fiducial pixel,
we can obtain real object fluxes from the $obj$ data.

After calculating the $obj$ data for all pixels, we made a map.
To fully utilize
the information of the
raster
positions and the optical distortion,
we used sub-pixels of 0.6 arcsec to create a map from the $obj$ data
(Appendix~\ref{sect:map}).
After some iterations to refine
the responsivity corrections with the cosmic ray and object masks
and the offsets among revolutions,
the final map was created.
Fig.~\ref{fig:deep} shows the final map
in the form of signal-to-noise ratio (S/N).
The $obj$ signals were co-added
with weights of their inverse variances
which were calculated by tracing the error propagation from the $dsub$ stage
(Appendix~\ref{sect:mad}).
Here we normalized standard deviations in sub-pixels in the coaddition
so that the central part of the histogram of the S/Ns in sub-pixels
should resemble a Gaussian distribution with a sigma of one,
because the noises in
sub-pixels are correlated. There was no indication of a strong source 
confusion in
the histogram, supporting the use of a Gaussian fit to the histogram
(Appendix~\ref{sect:noise}).

\begin{figure*}
   \centering
   \framebox[18cm]{1850\_f2.jpg}
   \caption{
The final map coadding all the data.
North is top and east to the left in J2000.
The center is (RA, Dec) $=$
($13^\mathrm{h}\,12^\mathrm{m}\,$26\fs0,
42\degr\,44\arcmin\,24\farcs8).
The map shows signal-to-noise ratio (S/N) per 0.6 arcsec sub-pixel.
The signal is an average,
weighted by the inverse of the assigned variance,
and the noise is a normalized standard deviation
(Sect.~\ref{sect:image}).
   }
   \label{fig:deep}
\end{figure*}


\section{Object extraction}
\label{sect:obj_ex}

\subsection{Determination of the detection parameters}
\label{sect:det_par}

For extracting sources, the SExtractor program
(Bertin \& Arnouts \cite{BA96})
was applied to Fig.~\ref{fig:deep}, the final S/N map.
The number of detected sources depends on three parameters:
the surface brightness threshold,
the number of 0.6 arcsec sub-pixels above the threshold,
and the width of the Gaussian filter used to convolve the map.

\begin{figure}
   \centering
   \resizebox{\hsize}{!}{\includegraphics{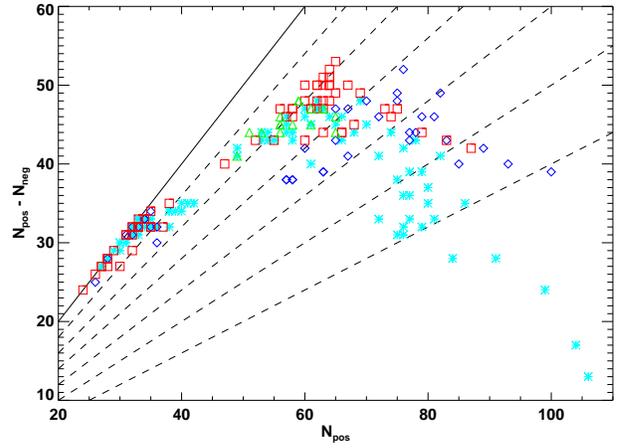}}
   \caption{
A search for the detection parameters
(Sect.~\ref{sect:det_par}).
The parameters for source detections
are: the surface brightness threshold,
the number of 0.6 arcsec sub-pixels above the threshold,
and the FWHM of the convolution Gaussian filter.
Diamonds denote a Gaussian FWHM of 3 sub-pixels,
asterisks 4 sub-pixels,
squares 5 sub-pixels,
and triangles 6 sub-pixels.
A peak in N$_\mathrm{pos}-$N$_\mathrm{neg}$
moves toward lower N$_\mathrm{pos}$ with the larger FWHM,
which corresponds to the suppression of noise detections.
The lines indicate the detection reliability
(N$_\mathrm{pos}-$N$_\mathrm{neg}$)$/$N$_\mathrm{pos} = \mathrm{100\,\%}$
(solid line),
90, 80, 70, 60, 50, and 40\,\% (dashed lines from the left to the right).
   }
   \label{fig:pos_neg}
\end{figure}

The detection parameters were determined
by comparing
numbers of detected sources on the S/N map
with those on the negative S/N map
(N$_\mathrm{pos}$ and N$_\mathrm{neg}$, respectively).
The negative S/N map was made by multiplying the S/N map by $-1$.
Any detections made on the negative map are spurious,
and no such sources should be detected (i.e. N$_\mathrm{neg}=0$)
where S/N is very high.
The source extractions on the positive and negative maps
were performed for more than 200 sets of the detection parameter values.
The results are shown
as a plot of N$_\mathrm{pos}-$N$_\mathrm{neg}$ against N$_\mathrm{pos}$
in Fig.~\ref{fig:pos_neg}.

For a perfect Gaussian noise field with positive source populations,
N$_\mathrm{pos}-$N$_\mathrm{neg}$ should increase with N$_\mathrm{pos}$
up to a certain
level and then flatten out at some N$_\mathrm{pos}$,
where the noise begins to be detected.  In fact,
Fig.~\ref{fig:pos_neg} shows a declining trend
where the flattening is expected.
This indicates that
the number of noise sources are somewhat larger in the negative side.
One explanation may be due to our technique for responsivity correction
(Sect.~\ref{sect:image}).
A failure in masking an object or a cosmic ray
will result in two dips in the neighboring data.
Such dips are actually identified around some bright sources in the final map
(Fig.~\ref{fig:deep}).
We believe that at least three dips are related to sources \#54, \#57, and \#22
(Table~\ref{tab:cat}).
Thus, their detection significance would be somewhat higher than the 
nominal values.

We chose the detection parameters
which gave the maximum of N$_\mathrm{pos}-$N$_\mathrm{neg}$:
1.04 in the normalized S/N in sub-pixels for the surface brightness threshold,
119 sub-pixels above the threshold,
and 5 sub-pixels for the width of the Gaussian.
This parameter set resulted in
N$_\mathrm{pos}=65$ and N$_\mathrm{neg}=12$.
These values give a nominal detection reliability of 82\,\%
for the whole sample.
However,
the true detection reliability would be somewhat higher
as indicated by the decreasing trend in Fig.~\ref{fig:pos_neg}.

\subsection{Simulations of the detection procedure}
\label{sect:simu}

\subsubsection{Detection completeness}
\label{sect:d_rate}

Monte Carlo simulations were performed
to evaluate the completeness of the source detection.
For each simulation,
a single test source was added to the real data,
the S/N map (Fig.~\ref{fig:deep}).
The profile of the test sources was fixed
as a two dimensional Gaussian function.
Its FWHM was determined by the real data;
the intensity weighted average
of the two smallest FWHMs among the ten most significant detections.
It was 12 sub-pixels or 7.2 arcsec.

Since we used the S/N map in the detection,
we assigned S/Ns to the test sources with the random signal values.
The noise was calculated for the 7.2 arcsec beam
as a quadratic sum of the sub-pixel noises in that beam.
This noise should be re-scaled,
because the sub-pixel noises are not independent.
The scaling factor was determined
so that the histogram of the S/Ns in the 7.2 arcsec beam
calculated over the entire map
should be a Gaussian with a sigma of one.
Thus, the noise includes all contributions from real sources
(Barger et~al. \cite{BCS99}).
The S/Ns were then determined with these signal and noise values.

\begin{figure}
   \centering
   \resizebox{\hsize}{!}{\includegraphics{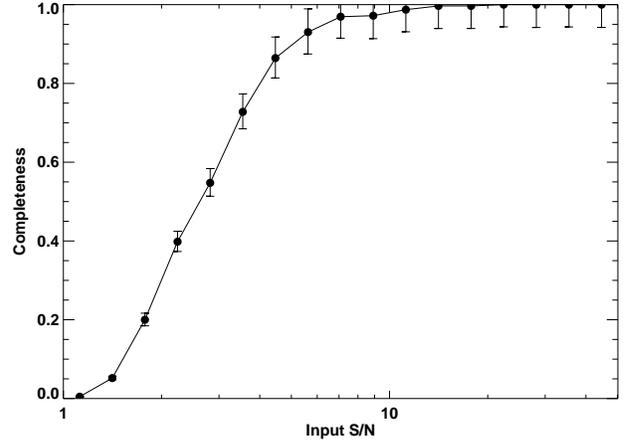}}
   \caption{
The completeness of the detection as a function of input S/N
(Sect.~\ref{sect:d_rate}).
A single test source was added to the S/N map (Fig.~\ref{fig:deep})
to evaluate the completeness of the detection procedure.
Almost 50,000 sets of S/Ns and positions were used.
One sigma errors were estimated assuming Poisson statistics
(Gehrels \cite{G86}).
   }
   \label{fig:d_rate}
\end{figure}

The simulations were done
with almost 50,000 sets of random S/Ns and positions for test sources.
As we used the real data,
we set some conditions for detections;
test sources should be detected within a distance of the FWHM
and the excess of the measured fluxes over the assigned ones
should be less than the assigned fluxes.
In the case that a test source was added near a certain real source,
some offset would be expected.
However, if the offset becomes larger than the FWHM,
this detection should be due to the pre-existing real source.
Similarly,
if the excess of the measured flux becomes larger than the assigned flux,
such a detection should be related to pre-existing real source contributions.
The result of the simulations is shown in Fig.~\ref{fig:d_rate}.
One sigma errors were estimated as Poissonian
(Gehrels \cite{G86}).
The completeness of the detection
is about 50\,\% and 80\,\% at assigned S/Ns of 2.5 and 4, respectively.

\subsubsection{Positional errors}

\begin{figure}
   \centering
   \resizebox{\hsize}{!}{\includegraphics{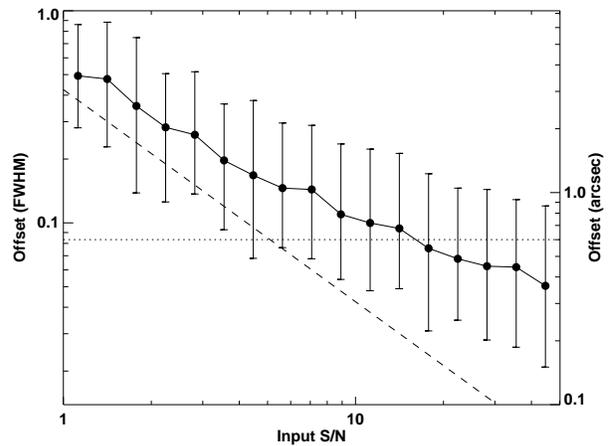}}
   \caption{
The offsets of the output coordinates from the input ones.
They are shown as a function of assigned S/Ns of the test sources.
One sigma errors were estimated with eq.~(\ref{eq:mad}).
Upper and lower sigma values were derived separately.
The dashed line denotes
the statistical formula for positional errors
$\sigma_\mathrm{pos}=\mbox{FWHM}/\sqrt{8\,\ln 2}/\mbox{(S/N)}$
(Condon \cite{C97}).
The size of sub-pixels in our map (0.6 arcsec) is also shown
(dotted line).
   }
   \label{fig:e_dist}
\end{figure}

Fig.~\ref{fig:e_dist} presents the simulation results
for the offsets of the measured coordinates from the assigned ones.
The offset is larger at lower S/N,
and is larger than the expected values given by the formula
$\sigma_\mathrm{pos}=\mbox{FWHM}/\sqrt{8\,\ln 2}/\mbox{(S/N)}$
(dashed line; Condon \cite{C97}).
A similar formula was used for a submillimeter map
with a high source density of 1/11 sources per beam
(Hughes et~al. \cite{HSD+98}).
Our map also has a high source density,
but 1/30 sources per beam.
We suggest that the deviation from the theoretical prediction
is related to the high source density
and also to the finite pixel size (dotted line) at high S/N.

\subsubsection{Flux errors and biases}
\label{sect:e_flux}

\begin{figure}
   \centering
   \resizebox{\hsize}{!}{\includegraphics{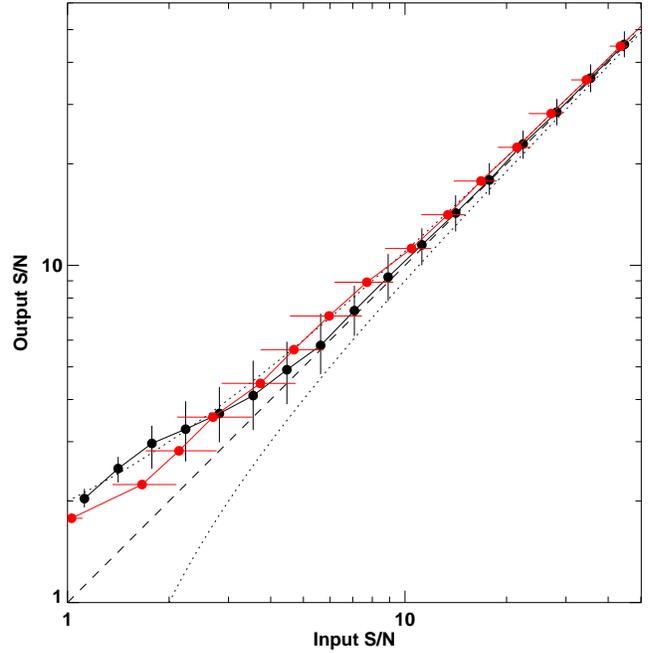}}
   \caption{
The output S/N of the detected source as a function of input S/N.
The individual simulation results are distributed
almost within the nominal one sigma limits
(dotted lines)
along the unity line
(dashed line).
Due to a combined effect
of the incompleteness and the source confusion,
the distribution is biased toward higher output S/Ns.
Means of the distribution and their one sigma errors are shown
with solid circles with vertical error bars.
For the correction of this flux biasing effect
in the direction from output to input,
the probability distribution over the diagram should be considered
(Sect.~\ref{sect:e_flux}).
It can be derived by convolving the simulation results
with the differential number counts
and their completeness along the input S/N axis.
The resulting relation from output S/Ns to input S/Ns
is presented with the red solid circles with horizontal error bars.
   }
   \label{fig:e_flux}
\end{figure}

Flux measurements always have errors.
This effect can be well understood
using the plot of assigned and measured fluxes
normalized by their noise values in Fig.~\ref{fig:e_flux}.
In this plot of input and output S/Ns,
we draw the unity line (dashed line)
and one sigma deviation lines
both for the positive and negative sides
(dotted lines).
The distribution of the simulation results on this plot
is fairly well represented by these two dotted lines.
Here we show simple moving averages
and their associated one sigma errors calculated from eq.~(\ref{eq:mad})
using solid circles with vertical error bars.
Large positive deviations from the unity line at low assigned S/Ns
are mainly due to the incompleteness of detections.
Almost no sources are detected with measured fluxes less than two sigma.
Even at relatively high S/Ns,
there are slight positive deviations from the unity line.
This could be due to effects of source confusion.

It should be noted that
this simple relation from input S/Ns to output S/Ns
cannot be applied in the opposite direction:
from measured fluxes to their true fluxes.
In this case one must take into account the probability distribution
of detected sources over this plot.
Such a distribution can be derived
by convolving the simulation results on this plot
with the probability distribution along the horizontal axis (true sources),
i.e., differential number counts,
and also the completeness of detection (Fig.~\ref{fig:d_rate}).

For the convolution calculation,
we introduced fine 0.03 dex meshes over the plot
in accord with the steep slope of the number counts
(Sect.~\ref{sect:nc}).
All we needed here was relative weights of the meshes
along the horizontal axis at each vertical position.
We assumed that the number counts were expressed with a single power.
In this case, the relative weights were determined by relative flux levels.
Thus, we could use the S/N horizontal axis.
We then applied the completeness values to these weights.
Convolving the final weights with the simulation results,
the transformation from measured S/Ns to statistically true S/Ns
was derived and is shown as the red solid circles with horizontal error bars
in Fig.~\ref{fig:e_flux}.
Note that the assumed power of the differential number counts
should be derived only after the correction of this flux bias.
Thus, some iterations were needed here.

Since the noises for the input and output S/Ns
are measured on the noise map and are thus irrelevant to the simulations,
this gives the relation between input and output fluxes.
Because the slope in number counts is negative,
fainter sources (less significant sources in this plot) are more numerous.
As a result, the true flux bias (red circles)
is larger than the simple relation (black circles)
where the completeness is almost maintained.
Beyond the point where the completeness begins to steeply decline to zero
at the low input S/Ns or faint fluxes,
the large deviations from the unity line seen in the simple relation
begin to be suppressed.

Flux biases for the detected sources
are listed in Table~\ref{tab:cat};
however, this correction is highly statistical.
For example, if a particular source is known to be isolated,
the correction derived here would be somewhat overestimated
since the effect of source confusion would be smaller for that source.

\section{The catalog}
\label{sect:cat}

Table~\ref{tab:cat} lists the
properties of all the 65 sources
detected with the detection parameters
described in Sect.~\ref{sect:det_par}.
Here we explain
the absolute calibration of the coordinates and the fluxes.

\begin{table*}
   \centering
   \caption[]{
   The catalog of ISOCAM LW2 sources in the SSA13 field.
     Column 1 is the identification number.
     Columns 2 and 3 are the Right Ascension in hours, minutes, seconds
	and the declination in degrees, arcmins, arcsecs in the J2000 system.
     Column 4 is the total astrometric error (one sigma) in arcsec.
	It is a quadratic sum of the absolute astrometric error
	and the relative astrometric error calculated with 0.2 dex 
bins in Fig.~\ref{fig:e_dist}.
     Column 5 is the normalized noise for the 7.2 arcsec diameter 
aperture at the source position.
	For its deviation, see Sect.~\ref{sect:d_rate}.
	For the correction from instrumental values,
	a sensitivity parameter of 2.32\,ADU/G/s/mJy is used
	(Blommaert \cite{B98}).
     Column 6 is the LW2 band (5--8.5\,$\mu$m) flux in $\mu$Jy.
	This is the total flux with all the flux corrections applied:
	the flux bias (Sect.~\ref{sect:e_flux}), aperture, and 
processing corrections
	(Sect.~\ref{sect:phot}).
	Taking account of the noise in all these corrections,
	their one sigma upper and lower errors are derived.
	SEDs are assumed to be flat
	($\nu f_{\nu} = \mathrm{constant}$).
     Column 7 is the flux bias defined as a ratio of true flux over 
measured one.
         Corresponding one sigma upper and lower limits are derived 
from Fig.~{\ref{fig:e_flux}}.
     Column 8 is the completeness of the detection
	derived with 0.2 dex bins in Fig.~{\ref{fig:d_rate}}.
     }
   \label{tab:cat}
   \catcode`?=\active \def?{\phantom{0}}
   \renewcommand{\arraystretch}{1.2}
   \begin{tabular}
     {rcccclcc}
       \hline
       \noalign{\smallskip}
                  Name &             RA &           Dec & 
$\Delta$r &         Noise &   6.7\,$\mu$m & Flux bias & Completeness 
\\
                       &    \multicolumn{2}{c}{(J2000)} & 
(arcsec) & ($\mu$Jy/beam)&    ?($\mu$Jy) &           &              \\
       \noalign{\smallskip}
       \hline
       \noalign{\smallskip}
       \input{1850_t3}
       \noalign{\smallskip}
       \hline
   \end{tabular}
\end{table*}

\begin{table*}[htbp]
   \centering
   \vspace{1mm}\leftline{{\bf Table \ref{tab:cat}.} Continued.}\vspace{4mm}
   \catcode`?=\active \def?{\phantom{0}}
   \renewcommand{\arraystretch}{1.2}
   \begin{tabular}
     {rcccclcc}
       \hline
       \noalign{\smallskip}
                  Name &             RA &           Dec & 
$\Delta$r &         Noise &   6.7\,$\mu$m & Flux bias & Completeness 
\\
                       &    \multicolumn{2}{c}{(J2000)} & 
(arcsec) &($\mu$Jy/beam) &    ?($\mu$Jy) &           &              \\
       \noalign{\smallskip}
       \hline
       \noalign{\smallskip}
       \input{1850_t3+}
       \noalign{\smallskip}
       \hline
   \end{tabular}
\end{table*}

\subsection{Astrometry -- coordinates}
\label{sect:astrom}

The coordinates in the catalog
were calibrated with the USNO A2.0 catalog
(Monet et~al. \cite{MBC+98})
extracted at the VizieR database
(Ochsenbein \cite{O97}).
Because very few LW2 sources were identified in the USNO catalog,
the calibration was performed in two steps.
The USNO catalog was first used to establish the coordinate system
in the $I$ band image that was taken from the ground
(Cowie et~al. \cite{CSH+96}).
The astrometric solution with 7 sources in the $I$ band image
shows a mean deviation of 0\farcs25.
The $I$ band image was then compared with the LW2 image.
Excluding confused sources,
12 sources in the LW2 band were used to derive the astrometric solution,
giving a mean deviation of 0\farcs75.
Then the absolute astrometric error
of the coordinate system in the LW2 image is 0\farcs79.
The major cause of this error is the combined effect of
the significant undersampling of the instrumental beam with 6 arcsec pixels
and the transients in the LW array
(Sect.~\ref{sect:phot}).
The total astrometric error for an individual source
is calculated as a quadratic sum of the absolute astrometric error
and the relative astrometric error determined with Fig.~\ref{fig:e_dist}.
For sources with high completeness,
this total astrometric error is conservative,
because the effects of finite pixels   have been accounted for twice.

\begin{figure}
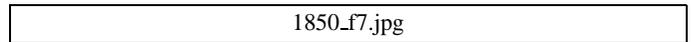

   \centering
   \framebox[9cm]{1850\_f7.jpg}
   \caption{
LW2 S/N contours superimposed on the $I$ band image.
The image registration is performed
using the astrometric solution obtained in Sect.~\ref{sect:astrom}.
The contour levels are set to 3, 4, 6, 10 and 18 sigmas in sub-pixels
(Fig.~\ref{fig:deep}).
   }
   \label{fig:i+lw2}
\end{figure}

Fig.~\ref{fig:i+lw2} shows
the LW2 S/N contours superimposed on the $I$ band image
with the astrometric solution derived above.
Many bright $I$ band sources are easily identified
with LW2 sources.
We counted numbers of bright ($I<23.7$) sources
for high and low completeness LW2 samples; there are
35 LW2 sources with completeness larger than 0.5
and 30 sources with lesser completeness.
Within a distance of 1.5 times the total astrometric error
from the measured coordinates,
28 and 24 $I$ band sources were found.
The surface density of $I<23.7$ sources suggests that
there would be 0.51 and 1.2 chance associations in the searched areas.
Subtraction of these numbers gives
rates of finding $I<23.7$ sources as 79 and 76\,\%
for the high and low completeness samples respectively.
This small decrease in the rate
indicates the high reliability of the catalog
even for sources with low completeness.

\subsection{Photometry -- flux}
\label{sect:phot}

The flux in the catalog is calculated statistically
to be a total flux for a single point-like source.
The total flux $F_\mathrm{total}$ is obtained with the formula
$F_\mathrm{total}=F_\mathrm{beam} \times 2.0 \, / \, 0.52$.
For each detected position,
an aperture flux with 7.2 arcsec diameter is measured in units of ADU/G/s
on the signal map.
This flux is corrected using the flux bias determined in 
Sect.~\ref{sect:e_flux}.
This gives $F_\mathrm{beam}$.
The two factors, 2.0 and 0.52,
represent aperture and processing corrections
which will be explained below.
The $F_\mathrm{total}$ in units of ADU/G/s
is then converted with a sensitivity parameter 2.32\,ADU/G/s/mJy
(Blommaert \cite{B98}).

The aperture correction was determined using the real data.
Growth curves of the detected sources
become noisy at diameters larger than 18 arcsec,
but the dispersion was almost a minimum at 7.2 arcsec.
Thus, the aperture correction from a 7.2 arcsec flux to an 18 arcsec flux
was adopted.
Similar to 7.2 arcsec flux measurements (Sect.~\ref{sect:e_flux}),
there would be flux biases in 18 arcsec flux measurements.
We therefore performed simulations for the 18 arcsec measurements
similar to those for the 7.2 arcsec measurements,
and corrected the flux bias in the 18 arcsec measurements.
The aperture correction factors were then obtained
using the bias corrected 7.2 and 18 arcsec flux measurements for the 
8 brightest sources,
which resulted in a median of 2.0 with a one sigma error of 9.8\,\%
(eq.~(\ref{eq:mad})).
At this stage, bias-corrected 18 arcsec fluxes were derived.

Next, the processing correction factor was derived
simulating all the image processing steps
by independent simulations different from those in Sect.~\ref{sect:simu}.
This processing correction is needed
not only to translate the bias-corrected 18 arcsec fluxes to total ones
but also to assess any effects in the image processing.
Such processing effects are usually canceled out
by a comparison with observational data for external calibration sources
that have been processed in the same manner.  However, in this case
no such data exist
due to the specification of the ISOCAM observations.

\begin{figure}
   \centering
   \resizebox{\hsize}{!}{\includegraphics{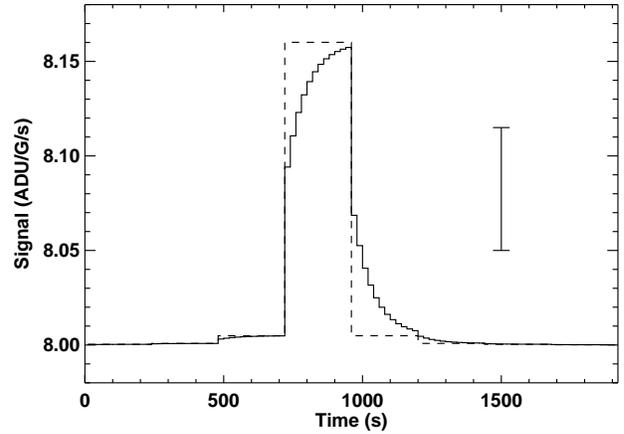}}
   \caption{
An example of the transient effect of the ISOCAM LW array.
The dashed line denotes
the signal from an ideal detector free from the transient effect.
A point source is assumed to pass over the center of the pixel
with raster movement having a step size of 7 arcsec.
The solid line represents
the expected signal with the transient formula
by Coulais \& Abergel (\cite{CA00}).
This figure assumes
an instantaneous response of $\beta = 0.51$,
a time constant of $\lambda = 560$\,ADU/G,
a background flux of 8\,ADU/G/s,
and a source flux of 100\,$\mu$Jy.
The vertical bar in the panel
is a typical one sigma noise for each exposure.
This implies that the transient correction for faint sources
would only be performed statistically on the final product.
   }
   \label{fig:trans}
\end{figure}

We therefore explain here the simulations used to derive
the processing correction
factor. At first, we needed to prepare simulated raw data.
A total flux was assigned to a simulated source
using the sensitivity parameter, and
the source was put at a random position on the sky
with the theoretical point spread function
(PSF; Okumura \cite{O98}).
It was then reprojected on the LW array
at every exposure position
using the satellite coordinates at every time step
and the measured offsets among revolutions.
The optical distortion coefficients of the LW array
(Aussel \cite{A98})
were also taken into account.
The derived source signals were modified
with the formula of the transient
(eq.~(6) in Coulais \& Abergel \cite{CA00})
using the measured background flux
(\textit{cf.} Fig.~\ref{fig:trans}).
Then, the measured responsivity drifts
and dark signals based on the dark model were applied.
Finally, we added the real ISOCAM raw data to these simulated raw data
to simulate noise.
This approach was mandatory because the noise properties in the ISOCAM data
had not been properly understood.

Next we applied the same image processing steps to these simulated raw data.
The PSFs of simulated sources on the final map
were found to be almost one arcsec sharper than those of real sources.
This suggests that the technique for making simulated raw data was not perfect.
One of the reasons for the sharper PSFs of the simulated sources
could be errors in the measured offsets among revolutions.
If the measured offsets were different from real values,
PSFs of real sources would become broader artificially,
while such effects cannot be involved in simulated sources.
Actually, the magnitude of the differences in PSF widths
is almost comparable to the total astrometric errors measured for real sources.
There could be other possibilities such as
errors in the optical distortion coefficients,
theoretical PSFs which are too sharp,
and ignoring cross-talk in the transient formula.

In any case, the
effects on the correction factor to the total flux
due to this discrepancy in the PSF shape
will be minimized for the large 18 arcsec aperture.
We found that
the bias-corrected 18 arcsec fluxes for the simulated sources
corresponded to $52^{+6.1}_{-4.7}\,\%$ of their total fluxes.
There was no meaningful correlation
between the correction factor and the input flux level.
One of the main reasons for the large deviation from 100\,\%
may be our technique for responsivity correction.
In this process,
non-stabilized source signals are divided
by responsivities which are too high due to transient remnants
({\it cf.} Fig.~\ref{fig:trans}).

\section{Number counts at 6.7\,$\mu$m}
\label{sect:count}

\subsection{Galaxy number counts}
\label{sect:nc}

To derive galaxy counts,
we have to identify stars in the catalog.
We adopted SExtractor stellarity indices
(Bertin \& Arnouts \cite{BA96})
measured with the ground-based $B$ and $I$ band images
(\textit{cf.} Fig.~\ref{fig:i+lw2}).
The stellarity indices were calibrated
with the $HST$ $I$ band image having a partial coverage of the field
(Cowie et~al. \cite{CHS95a}).
Two bright ($>100\,\mu$Jy) LW2 sources,
\#22 and \#64 in the catalog,
were then categorized as stars
based on their high stellarity indices.
At galactic latitude $|b|=50\degr$,
Franceschini et~al. (\cite{FTM+91}) estimated
that galaxies should dominate the 6.7\,$\mu$m source counts
below 90\,$\mu$Jy.
At this flux level,
star counts should be 0.15 arcmin$^{-2}$,
which is similar to our result of
two stars in our survey area (16 arcmin$^{2}$).
In the following discussion,
all the sources in the catalog are assumed to be galaxies
except \#22 and \#64.

The number count is the number of sources in an unit solid angle
with a certain flux constraint.
It can be obtained from an observation
as a sum of
inverse of maximum solid angles
in which the sources could be detected;
\begin{equation}
	N(S) = \sum_i 
\left(\frac{1}{\Omega_i}\right),	\label{eq:nc}
\end{equation}
where $N(S)$ denotes a differential number count in a flux bin at flux $S$,
$i$ is a running number for detected sources in the corresponding flux bin,
$\Omega_i$ is a maximum solid angle
having a sensitivity to detect the source $i$.
Integral number counts $N(>S)$ will be obtained by summing up $N(S)$.
In order to avoid unwanted effects due to negative sources,
we restrict the summation to the significance level
that could exclude negative sources.
This level corresponds to a completeness of 56\,\%.
Then the least significant source in the summation
has a S/N of 4.3 in the map coaddition.

\begin{figure}
   \centering
   \resizebox{\hsize}{!}{\includegraphics{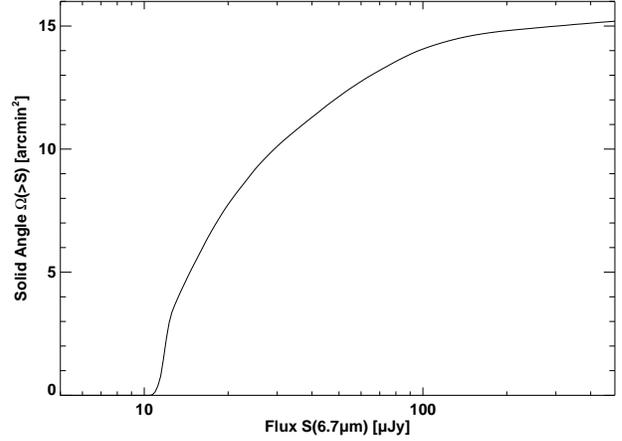}}
   \caption{
Survey solid angles as a function of 6.7\,$\mu$m flux.
Completeness of the detection is taken into account
down to 56\,\%. Below this limit, the completeness is
set to zero.
   }
   \label{fig:omega}
\end{figure}

Any incompleteness requires a calculation of effective solid angles.
As the noise level is not uniform in our map,
the effective solid angle $\Omega$
should be obtained as an integral;
\begin{equation}
	\Omega(>S) = \int \gamma(S/N) d\Omega(N),	\label{eq:omega}
\end{equation}
where $\gamma$ denotes a completeness as a function of $S/N$.
The map is divided into small pieces $d\Omega(N)$ giving the same noise level $N$.
We convolve the non-uniform noise map with a 7.2 arcsec diameter aperture
and relate them with the photometry of the detected sources
using the aperture and processing correction factors
(Sect.~\ref{sect:phot}).
To go well with the restriction of the summation,
$\gamma$ is set to zero for $\gamma < 0.56$.
The derived effective solid angles $\Omega(>S)$
are shown in Fig.~\ref{fig:omega} as a function of flux.

Then, integral galaxy number counts $N(>S)$ at 6.7\,$\mu$m
are obtained as listed in Table~\ref{tab:nc}.
The integral number counts climb up to $1.3\,\times\,10^4$ deg$^{-2}$
at the faintest limit.
The counts can be approximated with a single power $-1.6$
between 13\,$\mu$Jy and 130\,$\mu$Jy as
\begin{equation}
	N(>S)
	= 2.1^{+0.5}_{-0.4} \times 10^3\;\mathrm{[deg}^{-2}\mathrm{]}
	\left(\frac{S}{40\;\mathrm{[}\mu\mathrm{Jy]}}\right)^{-1.6 \pm 0.3}.
\end{equation}
As the integral counts are not independent of each other,
we also list the differential galaxy number counts $ N(S) $ 
at 6.7\,$\mu$m
in Table~\ref{tab:nc}.

\begin{figure*}
   \centering
   \resizebox{\hsize}{!}{\includegraphics{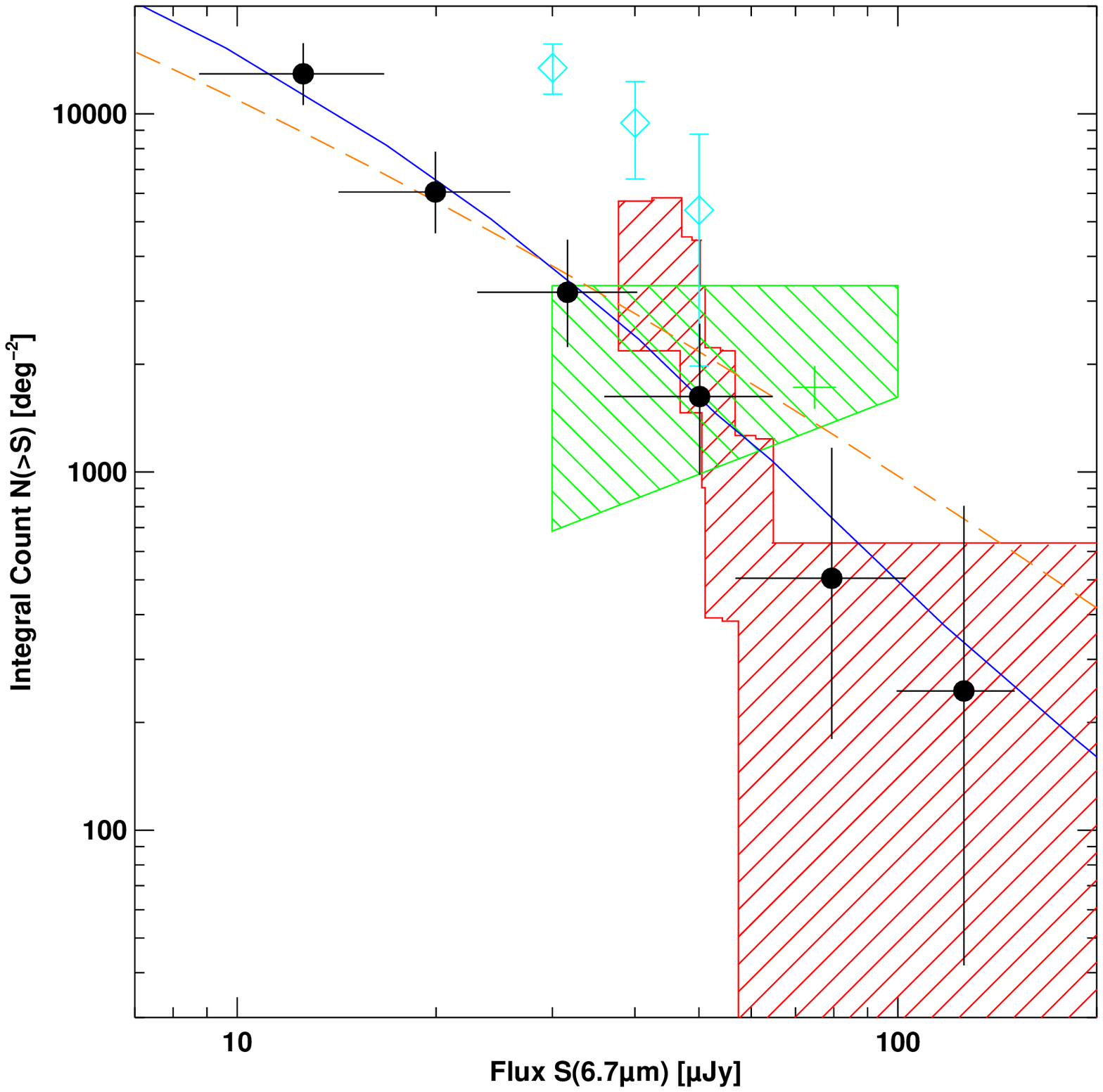}}
   \caption{
The integral galaxy number counts at 6.7\,$\mu$m.
Our counts in the SSA13 field are marked with black filled circles
with corresponding error bars (Sect.~\ref{sect:nc}).
The initial results and the re-evaluation of the HDF data
are shown as red hatches
(Oliver et~al. \cite{OGF+97})
and green hatches and a cross
(Aussel et~al. \cite{ACE+99}).
The counts for field galaxies
through the cluster gravitational lens A2390
are marked as cyan diamonds with their error bars
(Altieri et~al. \cite{AMK+99}).
Model predictions are overlaid for two models.
The blue line
represents the Franceschini et~al.\ (\cite{FAB+97}) model
and the dashed orange line is the Rowan-Robinson (\cite{R01}) model.
   }
   \label{fig:inc}
\end{figure*}

The integral galaxy number counts are shown in Fig.~\ref{fig:inc}
with error bars.
The vertical bars are derived with the Poisson statistics
(Gehrels \cite{G86}).
The horizontal bars were obtained
applying mean logarithmic flux error factors for the detected sources
counted in those flux bins.
The error ranges in flux becomes larger at fainter flux bins.

\begin{table*}
   \centering
   \caption[]{
The galaxy number counts at 6.7\,$\mu$m.
$N(>S)$ are integral counts and $N(S) = \Delta N(>S) / \Delta \log S$ are differential counts.
N$_\mathrm{pos}(>S)$
shows a number of the detected sources with a flux larger than $S$,
excluding the two stars.
N$_\mathrm{neg}(>S)$
is a number of the detected sources in the negative map.
N$_\mathrm{pos}(S)$ and N$_\mathrm{neg}(S)$
are numbers of the detected sources in the corresponding flux bins.
     }
   \label{tab:nc}
   \catcode`?=\active \def?{\phantom{0}}
   \begin{tabular}{lccccccc}
       \hline
       \noalign{\smallskip}
$\log S$         & $S$       & N$_\mathrm{pos}(>S)$ & 
N$_\mathrm{pos}(S)$ & N$_\mathrm{neg}(>S)$ & N$_\mathrm{neg}(S)$ & 
$N(>S)$      & $N(S)$ \\
$\log$ ($\mu$Jy) & ($\mu$Jy) &                      & 
&                      &                     & (deg$^{-2}$) & 
(deg$^{-2}$)      \\
       \noalign{\smallskip}
       \hline
       \noalign{\smallskip}
2.3 & 200 & ?0 & & 0 & & $0.0$ & \\
2.2 & 160 & & ?1 & &  0 & & $2.4\,\times\,10^2$ \\
2.1 & 130 & ?1 & & 0 & & $2.4\,\times\,10^2$ & \\
2.0 & 100 & & ?1 & &  0 & & $2.6\,\times\,10^2$ \\
1.9 & ?79 & ?2 & & 0 & & $5.1\,\times\,10^2$ & \\
1.8 & ?63 & & ?4 & &  0 & & $1.1\,\times\,10^3$ \\
1.7 & ?50 & ?6 & & 0 & & $1.6\,\times\,10^3$ & \\
1.6 & ?40 & & ?5 & &  0 & & $1.6\,\times\,10^3$ \\
1.5 & ?32 & 11 & & 0 & & $3.2\,\times\,10^3$ & \\
1.4 & ?25 & & ?7 & &  0 & & $2.9\,\times\,10^3$ \\
1.3 & ?20 & 18 & & 0 & & $6.1\,\times\,10^3$ & \\
1.2 & ?16 & & 12 & &  0 & & $6.9\,\times\,10^3$ \\
1.1 & ?13 & 30 & & 0 & & $1.3\,\times\,10^4$ & \\
       \noalign{\smallskip}
       \hline
   \end{tabular}
\end{table*}

\subsection{Extragalactic background light}

Extragalactic background light (EBL) due to resolved galaxies
can be derived as a first moment of the number counts:
\begin{equation}
	\nu I_{\nu}(S) =
	c/6.7\,\mu\mathrm{m} \times
	\sum_i \left(\frac{S_i}{\Omega_i}\right), 
	\label{eq:ebl}
\end{equation}
where $\nu I_{\nu}(S)$ is the 6.7\,$\mu$m EBL,
$S_i$ is the 6.7\,$\mu$m flux of the source $i$,
and $c$ is the speed of light.
Integrating the extragalactic contributions
in the flux range of $\log S=1.1$--2.3 for our counts
(Table~\ref{tab:nc}),
we obtain the integrated 6.7\,$\mu$m light of 
$0.56^{+0.12}_{-0.10}$\,nW\,m$^{-2}$\,sr$^{-1}$.
The upper and lower errors are derived as in Sect.~\ref{sect:nc}.
It should be noted that
this integrated light is still a lower limit to the 6.7\,$\mu$m EBL,
because the slope of the number counts
is steeper than $-1$ even at the faint bins.
So the 6.7\,$\mu$m EBL would become larger
when integration to fainter flux levels becomes possible.

Recently, TeV $\gamma$-ray data were used to put constraints
on the EBL in the infrared. This is based on electron-positron
pair creation by photon collisions $\gamma_\mathrm{TeV}
+ \gamma_\mathrm{IR} \to \mathrm{e}^\mathrm{+} + \mathrm{e}^\mathrm{-}$.
Upper limit values on the infrared EBL are known to be sensitive to
the assumed shape of the infrared EBL.
If the infrared EBL had a local minimum around 5\,$\mu$m,
the $5\,\sigma$ upper limit could become as low as 2\,nW\,m$^{-2}$\,sr$^{-1}$
(Renault et~al. \cite{RBL+01}).
With a wide range of reasonable assumptions on the infrared EBL,
a strict upper limit of 4.7\,nW\,m$^{-2}$\,sr$^{-1}$
between 5 and 15\,$\mu$m was derived so far.
From the galaxy modeling side,
a value of 2.4\,nW\,m$^{-2}$\,sr$^{-1}$ at 6.7\,$\mu$m was predicted
with the assumed redshift of galaxy formation $z_\mathrm{F}=9$
(Franceschini et~al. \cite{FTM+91}).
Much more precise values of the EBL in this spectral range
would put a strong constraint on the major epoch of star formation
in galaxies
(Primack et~al. \cite{PBS+99}).

\subsection{Comparison with other 6.7\,$\mu$m counts}

Our counts go somewhat deeper than previously reported counts at 6.7\,$\mu$m
(Fig.~\ref{fig:inc}).
In the overlapping flux range,
our 6.7\,$\mu$m galaxy counts are almost consistent
with the HDF counts
(Oliver et~al. \cite{OGF+97};
Aussel et~al. \cite{ACE+99}).
While the recent HDF-S counts (Oliver et~al. \cite{OMC+02}, not shown)
are somewhat larger than the HDF and our counts,
the field galaxy counts lensed by the massive cluster A2390
(Altieri et~al. \cite{AMK+99})
are about four times larger than ours.

The excess in the A2390 counts
might be due to a failure in identification of cluster members.
There may be errors in gravitational lensing corrections in flux and 
solid angle.
Multiple lensed images might be counted as individual field galaxies.
These effects could become more significant
since
the target field is at the center of the cluster
and the surveyed area is more than two times smaller than ours.
Another reason might be contamination by stars.
The SSA13 field and the HDFs are located at high galactic latitudes:
$b \sim 74\degr$\ (SSA13), $b \sim 55\degr$\ (HDF), and $b \sim 
-49\degr$\ (HDF-S).
A2390 is located at low galactic latitude ($b \sim -28\degr$).
At $|b|=20\degr$, stars are expected to dominate the counts
even at a faint flux level of 3\,$\mu$Jy
(Franceschini et~al. \cite{FTM+91}).
Altieri et~al. (\cite{AMK+99})
stated that only two stars were identified among their 31 detections.
The
solid angles of the Altieri et~al. (\cite{AMK+99}) counts might be 
underestimated
by subtraction of the pixel areas occupied with brighter sources.
Aussel et~al. (\cite{ACE+99}) did this
and the nominal Aussel et~al. (\cite{ACE+99}) count
(green cross) is larger than
the Oliver et~al. (\cite{OGF+97}) counts in the HDF.

\subsection{Comparison with models}

There are essentially two approaches to predict galaxy number counts.
One is a parametric evolution approach,
which has been applied to IRAS 60\,$\mu$m galaxy counts
using IRAS 60\,$\mu$m local luminosity function and its population mix.
Normal spirals, starbursts, and AGNs
are assumed to have non-evolving empirically determined SEDs
but a parametric evolution in luminosity and/or density.
After the observations with ISO and SCUBA,
many authors have extended this modeling
mainly to explain ISO 15\,$\mu$m counts and SCUBA 850\,$\mu$m counts
(Xu et al. \cite{XLS+03} and references therein).
Some groups predict galaxy counts at 6.7\,$\mu$m as well.
Here we show the Rowan-Robinson (\cite{R01}) model
(dashed orange line in Fig.~\ref{fig:inc}),
which could be marginally consistent with our counts.
This model is known to  represent well bright 6.7\,$\mu$m counts.
This will be due to the following facts.
In the local universe,
the observing band of the ISOCAM LW2 (5--8.5\,$\mu$m) filter
matches the location of PAH and hot dust emission,
which could be strong in star forming galaxies.
Such low-redshift star-forming galaxies
are known to be traced in the far-infrared,
such as in the IRAS 60\,$\mu$m band.

Another approach is a stellar population synthesis method,
which is used to model optical and near-infrared galaxy counts
in searches for cosmological parameters.
E/S0 and spiral galaxies are assumed and their local luminosity functions
are adopted in the optical or near-infrared.
Their SEDs evolve with time as their stellar populations do.
Some include an AGN contribution as a parametric evolution population.
Franceschini et al. (\cite{FAB+97}) used this approach to model 
6.7\,$\mu$m counts
(blue line in Fig.~\ref{fig:inc}).
Although the observational counts are associated with large errors,
the consistency between this Franceschini et al. (\cite{FAB+97}) model
and our results is remarkable.
The steep slope comes from stellar emission of high redshift ($z>1$) 
E/S0 galaxies.
At this redshift, the ISOCAM LW2 band will start to select
near-infrared emission from stellar photospheres in galaxies,
especially in E/S0 galaxies
which are assumed to have formed most of their stars in the early epochs.
Moreover, stellar SEDs short-ward of 6.7\,$\mu$m are steep up to 1\,$\mu$m
and their brightening at high redshifts is also expected.
Because E/S0 galaxies lack PAH and dust emission in the local universe,
there is a negligible contribution from E/S0 galaxies in bright 
6.7\,$\mu$m counts.
This affects additional contributions from E/S0 galaxies
in this faint flux range.
In contrast with its good agreement with the faint 6.7\,$\mu$m counts, the
Franceschini et al. (\cite{FAB+97}) model is known to underpredict
bright 6.7\,$\mu$m counts, probably due to insufficient evolution of 
dusty galaxies
in the stellar population synthesis method.

In this respect,
a trial to combine these two approaches has been made  by Xu et al. 
(\cite{XLS+03}).
After constructing three parametric evolution models,
each of which is consistent with existing dusty galaxy counts,
they  added E/S0 galaxy counts to each of them.
The resulting counts have a  slope consistent with our counts,
but the counts themselves are much higher than ours, and in fact are
more consistent with the A2390 counts.
We think that their population of starbursts should be merged
with E/S0 galaxies at high redshifts.
Xu et al. (\cite{XLS+03}) themselves report
a synchronization of active evolutionary phases of E/S0s and 
starbursts at high redshifts ($z>1$).
A method of stellar population synthesis has now been extended to the 
far-infrared,
and it shows that SEDs of E/S0 galaxies can resemble those of starbursts
in their early evolutionary phases
(Silva et~al. \cite{SGBD98}).
Interestingly, Pearson (\cite{P01}) predicts
excessive 6.7\,$\mu$m counts similar to those of Xu et al. (\cite{XLS+03})
by adding a ULIG component to his parametric evolution model instead of an E/S0 component.

On the observational side,
we find all submillimeter sources in the SSA13 field
have SEDs consistent with high redshift ($z>1$) galaxies
having stellar masses comparable to those of typical E/S0s
(Sato et~al. \cite{SCK+02}).
At the shorter wavelengths,
a deficiency of high redshift ($z>1$) elliptical galaxies was reported
from optical observations with spectroscopic and photometric redshifts
(Franceschini et~al. \cite{FSF+98}).
The number of morphologically identified elliptical galaxies
at faint magnitudes ($I>24$)
was found to be smaller than the model predictions
(Driver et~al. \cite{DFC+98}).
Instead, the number of disturbed faint galaxies
was reported to be larger
and some of them were identified with high redshift galaxies
(Cowie et~al. \cite{CHS95a}, \cite{CHS95b}).
All these support the idea that
elliptical galaxies undergo dusty starbursts at high redshifts.
However, similar number count predictions can be reproduced
with a different set of SEDs and evolutionary parameters,
especially in a parametric evolution approach
as shown in the three parametric evolution models
presented by Xu et al. (\cite{XLS+03}).
Thus, a definite answer must await detailed investigations
of individual faint 6.7\,$\mu$m galaxies detected in this survey
(Sato et~al. \cite{SCK+03}).


\section{Summary}

A very deep mid-infrared survey has been conducted
to investigate evolution of field galaxies at high redshift.
The high galactic and high ecliptic latitude field
SSA13 has been imaged with the ISOCAM's broad band filter at 6.7\,$\mu$m.

Utilizing quite high redundancy with a total observing time of 23 hours,
we have developed a correction for responsivity drifts.
We also introduced several types of masking
and median noise estimates for data affected by outliers
(Appendix~\ref{sect:details}).
Many simulations have been carried out to properly account for
the characteristics of the image processing and source detection.
Within an area of 16 arcmin$^2$, 65 mid-infrared sources down to
6\,$\mu$Jy have been detected.
For the central 7 arcmin$^2$,
an 80\,\% completeness limit reaches 16\,$\mu$Jy.

The galaxy number counts have been derived
with statistical corrections.
The slope of the integral counts is
$-1.6$
between 13\,$\mu$Jy and 130\,$\mu$Jy
and the counts exceed $1\,\times\,10^{4}$ deg$^{-2}$ in the faint limit.
Our counts reach a limit about three times fainter than those in the HDFs.
Although more detailed investigations are necessary,
the steepness of the counts might indicate
that stellar emission of evolving E/S0 galaxies
has redshifted into the ISOCAM LW2 bands at this faint level.


\begin{acknowledgements}

YS would like to thank Minoru Freund and Mikako Matsuura
for their dedicated hardware provisions in the early phases.
We have received many useful comments from Leo Metcalfe,
the leader of ISOCAM Instrument Dedicated Team
during the ISO operational phase.
Leo Metcalfe and Bruno Altieri have suggested
the cosmic ray deglitching
based on the data pointing at the same sky position.
Discussions with Shinki Oyabu were very useful.
We are indebted to the first referee for the detailed suggestions
and the second referee for complementary and constructive comments
during the very long referring process.
We also thank the referee of a subsequent paper,
who gave us an essential comment on the incompleteness correction.
The analysis has been achieved with the IDL Astronomy Users Library
maintained by Wayne Landsman.
This research has
made use of NASA's Astrophysics Data System Bibliographic Services.
A part of this research has been supported by
JSPS Research Fellowships for Young Scientists.

\end{acknowledgements}


\appendix

\section{Details of the image processing}
\label{sect:details}

The image processing is summarized in Sect.~\ref{sect:image}.
Here we describe details of that processing;
creation of $raw$, $dsub$, $sresp$, $ofrac$, and $obj$ data,
object masking, cosmic ray masking,
and noise estimates with median absolute deviation.
Map creation and noise investigation are also described.

\subsection{Dark subtraction}

Although all the processing steps in Sect.~\ref{sect:image}
are considered for one detector pixel,
procedures in this subsection can be applied frame by frame.
The raw signal frame $raw$ in eq.~(\ref{eq:raw})
was derived as a difference of ISOCAM image frames
taken at the reset and the end of integration.
Subtraction of the dark image frame $b$ from the $raw$ frame
gave the dark-subtracted frame $dsub$.
The dark image frames were created with the dark model
invented and updated by Biviano et~al. (\cite{BSG+98})
and Roman \& Ott (\cite{RO99}).

\subsection{Median absolute deviation as a noise estimator}
\label{sect:mad}

The subsequent processing requires threshold values
to identify objects and cosmic rays
such as 3 sigma detection or 3 sigma rejection thresholds.
In order to obtain such thresholds,
one of the robust noise estimators --
median absolute deviation --
was adopted.
Median absolute deviation $\tau$ is defined
here
as
\begin{equation}
	\tau = \mathrm{med}(\mathrm{abs}(x_i - \mathrm{med}(x_i)))
   \label{eq:tau}
\end{equation}
for sample {$x_i$}.
Detector noise, the sum of readout and photon noises,
can be represented with the $\tau$,
because the median operations in eq.~(\ref{eq:tau})
are insensitive to outliers such as cosmic rays.
The $\tau$ was then scaled
to give it the statistical meaning of the Gaussian sigma $\sigma$ as
\begin{equation}
	\sigma_{\tau} = \tau / \tau_0,
   \label{eq:mad}
\end{equation}
where $\tau_0 = 0.67448975$.
The scaling factor $\tau_0$
is the $\tau$ for a sample with a Gaussian distribution with $\sigma=1$.

For the $dsub$ data,
$\sigma_{\tau}$ was calculated for each pixel in each revolution.
With these $\sigma_{\tau}$ as starting values,
all the error propagation was traced
in the subsequent arithmetic operations.

\subsection{Responsivity correction}
\label{sect:resp}

\subsubsection{Raster point masking}
\label{sect:raster}

When we are looking at blank sky,
detector signals are dominated by the zodiacal emission.
The zodiacal emission depends on the solar aspect angle
(Table~\ref{tab:rev}).
The temporal variation of the solar aspect angle
is almost negligible during observations lasting less than several hours.
Thus, the responsivity of the detector can be monitored
by using blank sky as a reference.

Because the responsivity drifts are major noise sources,
creating the response to sky $sresp$
is the most essential part of the image processing.
The $sresp$ data were derived from the dark-subtracted data $dsub$.
Because of the different sky levels among revolutions
(Table~\ref{tab:rev}),
this operation was performed separately
with the data taken in each individual revolution.

\begin{figure}
   \centering
   \resizebox{\hsize}{!}{\includegraphics{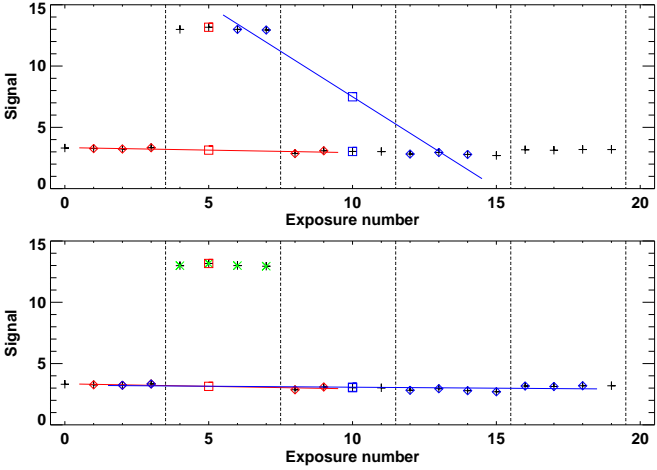}}
   \caption{
Schematic diagrams for the derivation of the response to the sky $sresp$.
The top panel shows the derivation with raster point masking.
The number of exposures per raster point is set to four
(i.e. N$_{\mathrm{exp}}=4$).
Plus signs show dark-subtracted $dsub$ signals of a detector pixel
during five raster point observations.
Vertical dotted lines show boundaries of raster points.
A $sresp$ value for exposure \#5 (a red square on a red line)
is derived with a line fitted to five $dsub$ samples (red diamonds).
In the bottom panel, object masks are applied.
Four $dsub$ samples at the second raster point
are masked out (green asterisks).
A $sresp$ value for exposure \#10 (a blue square on a blue line)
is then derived correctly
with a line fitted to nine unmasked $dsub$ samples (blue diamonds).
   }
   \label{fig:sresp}
\end{figure}

The top panel of Fig.~\ref{fig:sresp} shows a schematic diagram for the
$sresp$ creation.
A $sresp$ value is to be assigned to each exposure.
This value is evaluated
via a linear interpolation of
four or more $dsub$ samples in adjacent raster points.
For example,
a $sresp$ value for exposure \#5
(a red square on a red line)
is derived by fitting a straight line to
three $dsub$ samples at the first raster point (\#1, 2, and 3)
and two $dsub$ samples at the third raster point (\#8 and 9).
Similarly, a $sresp$ value for exposure
\#6 will be derived from five $dsub$ samples (\#2, 3, 8, 9, and 10).
Because no $dsub$ samples
in its own raster point is used to derive a $sresp$ value,
we call this technique \textit{raster point masking}.

In the neighborhood of exposures with contributions from objects,
however, the raster point masking gives higher $sresp$ values than 
the correct ones.
Such an example is shown for a $sresp$ value for exposure \#10
(a blue square on a blue line). In this setting, all the exposures
at the second raster point should be masked out, because the detector
is looking at a bright object at that raster point. With such masks,
$sresp$ values are then derived correctly. We will discuss such a
procedure just after explaining object masks below.

\subsubsection{Object masks}
\label{sect:objmask}

With the responsivity correction,
division of the $dsub$ data with the $sresp$ data
(eq.~(\ref{eq:ofrac})),
and the subsequent normalization
(eq.~(\ref{eq:obj})),
the $obj$ data were obtained.
They were calibrated to have zero flux for the sky.
Thus, any positive features should be objects.
The estimated noise is
around 30\,$\mu$Jy per pixel (1\,$\sigma$) at this stage.
Thus, almost no object could be identified in our $obj$ data
(Sect.~\ref{sect:phot}).

A low noise map must be derived
in order to determine the location of the objects.
All the $obj$ data taken in a single revolution
are then coadded (Appendix~\ref{sect:map}).
The resulting map is called a \textit{revolution map}.
The noise found in the revolution map
was more than ten times smaller
due to coaddition of a few hundred $obj$ data in a revolution.
At this stage,
many bright sources could be identified with a certain threshold.
Note that median values were used in the coaddition
to suppress contamination by cosmic rays.

The objects on the 2-dimensional (2-D) revolution map needed to be 
located in the 3-dimensional (3-D) data,
because the $sresp$ data are calculated with the 3-D $dsub$ data.
However, there is a complicated relation
between sub-pixels in the 2-D revolution map and pixels in the 3-D data,
due to raster slews, the optical distortion of ISOCAM
and the use of sub-pixels in a revolution map (Appendix~\ref{sect:map}).
Thus, the projection of the 3-D $obj$ data to the 2-D revolution map 
was reversed.
The resulting 3-D data are called reversely-projected object $robj$ data.
Any pixel in the 3-D $robj$ data can be compared with that in the 3-D 
$dsub$ data.
This transformation from 2-D to 3-D is designed
so that S/N in the 3-D $robj$ data is the same as in the 2-D revolution map.
A single S/N threshold was then used to mask pixels in the 3-D data 
looking at objects.

\subsubsection{Raster point masking with object masks}
\label{sect:rpmwom}

Now, we could apply the object masks to the $dsub$ data.
The raster point masking (Appendix~\ref{sect:raster}) was then repeated
on the object-masked $dsub$ data to refine the $sresp$ data.

The bottom panel of Fig.~\ref{fig:sresp}
shows a schematic diagram for the raster point masking with object masks.
All the $dsub$ data samples at the second raster point are 
object-masked (green asterisks).
A $sresp$ value for exposure \#10 (a blue square on a blue line)
is then derived from nine $dsub$ samples \#2, 3, 12, 13, 14, 15, 16, 
17, and 18 (blue diamonds).
Note that a $sresp$ value is estimated as the midpoint of the fitted line
and at least two samples are required at one side of the fitted region
to properly interpolate the responsivity drifts.

In order to avoid inadequate interpolation,
pixel history was regarded as discontinuous
at positions having more than $3\,\sigma$ deviation from the previous data.
Near the discontinuities,
the conditions of linear interpolation above can not be satisfied.
Any exposures for which $sresp$ values could not be defined
were not used for the subsequent processing
(Fig.~\ref{fig:history3}).

The raster point masking with object masks was repeated three times.
The object identification thresholds were decreased;
5, 4, and 3 sigmas in this order.
Note that no normalization of the noises in 6 arcsec pixels
has been applied.
At the final stage, the object masks occupy 3\,\% of the data.

\subsubsection{Dwelling time and timescale of the responsivity drifts}

Some discussion is warranted concerning an appropriate value for
the number of exposures per raster point;
N$_{\mathrm{exp}}=12$ for our case.
Suppose one pixel looked at an object at the $i^{\mathrm{th}}$ raster point
and it looked at sky at the adjacent raster points.
The raster point masking routine calculates the $sresp$ values
for the $i^{\mathrm{th}}$ raster point
by interpolating the $dsub$ data at the adjacent raster points.
If the $dsub$ data at the $i^{\mathrm{th}}$ raster point are
larger than the interpolated $sresp$ data,
this excess is recognized as an object.
Because the responsivity is drifting,
the noise caused in the interpolation becomes larger,
if the time interval increases
between the data at the $i^{\mathrm{th}}$ raster point
and the reference data at the adjacent raster points.

To evaluate this effect,
a trial map was created with an assumption
that the data were taken as if N$_{\mathrm{exp}}=1$
(Fig.~\ref{fig:clean}).
The object masks were created from the final map (Fig.~\ref{fig:deep})
with an object identification threshold of 2.5 sigma.
As expected,
fluctuation in the background became roughly four times smaller
than that in the final map.
This indicates that smaller N$_{\mathrm{exp}}(<12)$
would be preferable to correct the responsivity drifts.
Note that any objects below the identification threshold
will never have appeared in this trial map.

\begin{figure}
   \centering
   \framebox[9cm]{1850\_fa2.jpg}
   \caption{
A trial map constructed as if the data were taken with N$_{\mathrm{exp}}=1$.
The object masks were made from the final map
(Fig.~\ref{fig:deep})
with 2.5 sigma for the identification threshold parameter.
For the display,
the noises in sub-pixels are normalized to correct
the noise correlation in the map creation stage.
   }
   \label{fig:clean}
\end{figure}

\subsection{Cosmic ray masks -- deglitching}
\label{sect:crmask}

The $dsub$ data contain many spikes
(top panel of Fig.~\ref{fig:history3}).
Such spikes, usually called glitches,
were caused by cosmic ray impacts on the detector pixels.
Glitches are short term phenomena, lasting a couple of exposures,
and thus differ from the long term responsivity drifts.
The deglitching algorithm was developed
to remove glitch effects in our data with T$_{\mathrm{int}}=20$\,sec.
There are three masking steps
based on $dsub$, $obj$, and $robj$ data.

Suppose that a glitch lasts $N$ exposures in the $dsub$ data.
To remove this glitch with a time scale of $N$,
the running median routine should be applied
with a width of $2N+1$.
Any deviant data from the running medians
were then regarded as glitches.
The scale of glitches was typically $N=2$;
a cosmic ray impact and its trailing remnant.
A new glitch sometimes occurred
before the preceding glitch had faded out.
To take account of such overlaps,
the scale was set to $N=3$.
Thus, $2N+1=7$ was set for the width of the running median.
This cosmic ray masking routine was repeated
until no data deviated from the threshold.
Then, an additional masking was applied with a scale of $N=11$,
equivalent to the number of exposures at a raster point
(N$_{\mathrm{exp}}=12$).

This first masking step on the $dsub$ data
may mask bright objects.
Thus, once the $sresp$ data were derived
without contamination from glitches,
all the masks were canceled.
New masks were then created from the $obj$ data.
To preserve object fluxes in this step,
a median of all exposures at each raster point
was calculated as a reference value for masking.
For example,
a median of the $i^{\mathrm{th}}$ raster point
was compared with all the data samples at this raster point.
Any data samples which deviated from the threshold were masked out.

The reference values for the first and second masking steps
were derived from two dozen samples at most.
Many points of sky within our field
were observed more than 1000 times.
Thus, in the third masking step,
the low noise $robj$ data were utilized
(Appendix~\ref{sect:objmask}).
All data in the $obj$ data were compared with
their respective $robj$ data.
Any data were masked if they deviated from the threshold.

Because we iterated the processing to refine object masks
(Appendix~\ref{sect:rpmwom}),
the operation returned
to the initial cosmic ray masking routine with the $dsub$ data
after the $robj$ masking step.
In the second iteration and later,
only the scale $N=2$, instead of 3 and 11,
was used to deglitch the $dsub$ data.

Different deglitching thresholds were set
for cosmic ray impacts and their trailing remnants;
typically 3 and 2.5 sigmas, respectively.
These thresholds were adjusted
to result in zero background in a map (eq.~(\ref{eq:ofrac}))
and no loss in object fluxes.
For discrimination of objects and cosmic rays,
a standard deviation map and
a median absolute deviation map (eq.~(\ref{eq:mad}))
were compared.
Objects appear in both maps
while cosmic rays appear only in the standard deviation map.
In the $robj$ deglitching,
an additional margin for the thresholds was set
to go well with the median nature of the $robj$ data
(Fruchter \& Hook \cite{FH98}).
At the final stage,
cosmic ray masks occupy 23\,\% of the data.

\subsection{Sky brightness}
\label{sect:sky}

The fiducial sky flux $I^0_{\mathrm{sky}}$ falling onto the fiducial pixels
must be known to derive the object flux $obj$ (eq.~(\ref{eq:obj})).
The fiducial pixels are the central $12 \times 12$ detector pixels
that have negligible effects from the optical distortion and the vignetting
($\alpha=1$ in eq.~(\ref{eq:alpha})).
The sky flux was dominated by the zodiacal light
and the zodiacal light depends on the solar aspect angle
(Table~\ref{tab:rev}).
Thus, the sky flux must be measured for each revolution.
The $I^0_{\mathrm{sky}}$ was determined in two steps:
(1) a median value of each pixel in the $sresp$ data was calculated
to minimize the effect of the responsivity drifts, and then
(2) these median values of the central $12 \times 12$ pixels were averaged.
The fiducial sky flux for each revolution is given in Table~\ref{tab:rev},
after the conversion to the sky brightness.

\subsection{Map creation}
\label{sect:map}

There exists an offset between the image on the detector and
the actual satellite pointing.
This is due to jitter in the lens wheel repositioning in the camera.
The offsets due to the jitter vary from revolution to revolution.
To evaluate these offsets,
a revolution map was first created
from the $obj$ data taken in each revolution.
At first, we describe the creation of the revolution maps.

Information about the satellite position
and the optical distortion at the ISOCAM detector
were used to coadd the $obj$ data.
To take account of the rotation and the distortion
of the 6 arcsec detector pixels on the sky,
a map with 0.6 arcsec sub-pixels
was created aligned with the J2000 system.
The position and the orientation of the ISOCAM detector were obtained
based on three angles defining the satellite status; RA, Dec, and Roll.
Then, each detector pixel area on the sky map was defined
with positions of the four corners of that pixel.
The positions were derived
with the optical distortion correction
dependent on the lens position
(Aussel \cite{A98}).
Finally, the distortion corrected pixel was shrunk to its 85\,\%
($0.85^2$ in area)
and put into sub-pixels.
This shrink operation recovers some information lost
in the undersampling of the optical beam
(Fruchter \& Hook \cite{FH98}).

The pointing stability within T$_{\mathrm{int}}=20$\,sec
was measured as 0.34 arcsec (1\,$\sigma$) on the average.
Frames having displacements twice as large as this value
were regarded as having been taken during slews of the satellite.
Such frames were not used for the coaddition.
The jitter of the lens wheel gave rise to a
weakly illuminated area at some edges of the detector array.
Pixels in such areas show 3 times larger noise in the error 
propagation calculations.
They were masked out in the coaddition.
The fraction of detector area affected by weak illumination
depended on the lens position (Table~\ref{tab:rev});
3\,\% of a detector frame for the $left$ position
and 1\,\% for the $right$.

After the revolution maps were obtained,
positions of bright sources identified in different revolutions were compared.
Based on the relative offsets of 19 sources,
offsets among revolutions were derived (Table~\ref{tab:rev}).
With the correction of these offsets to the RA and Dec angles of the satellite,
the final map (Fig.~\ref{fig:deep})
was calculated in the same way to obtain revolution maps.
Here all the $obj$ data in all the revolutions were used.

Note that the relative offsets of the 19 sources
had a mean deviation of 0.13 pixel, 0.8 arcsec.
If such uncertainty is acceptable,
\emph{a priori} jitter correction can be performed.
Namely,
the offset in the M-direction is 1 pixel for the $left$ position
and $-1$ pixel for the $right$.
The offset in the raster N-direction is 0
(Table~\ref{tab:rev}).

\subsection{Unfolding noise contributions}
\label{sect:noise}

The noise in the 7.2 arcsec beam was
estimated to be typically 1.1\,$\mu$Jy/beam,
with a use of a histogram for the correct absolute scaling of the noise.
This noise includes detector, processing, and confusion noises.
Here we tried to unfold these noise sources.

For estimating a confusion noise,
we adopted an evolution model by Franceschini et~al.\ (\cite{FAB+97})
because of the small deviation of our counts from this model.
For an extension toward fainter fluxes,
we assumed a parameterized form for the Franceschini et~al.\ model;
\begin{equation}
	N(>S) = 
\frac{2\,N_0}{\left(\frac{S}{S_0}\right)^{{\alpha}_{1}}+\left(\frac{S}{S_0}\right)^{{\alpha}_{2}}}
\end{equation}
with a set of parameters $(\alpha_1,\alpha_2,S_0,N_0)
=(0.70,1.95,20\,\mu\mathrm{Jy},6.4 \times 10^3 \mathrm{deg}^{-2})$.
This formula represents the model quite well at fluxes less than 100\,$\mu$Jy.
We generated test sources following this parameterized relation down 
to 0.05\,$\mu$Jy
and distributed them randomly over the sky.
The fluctuation in the 7.2 arcsec beam over this map
was estimated with a histogram, giving 0.49\,$\mu$Jy/beam as a modal value.

Using this simulated data set,
we created raw data
and all the image processing was executed as in the processing simulations
in Sect.~\ref{sect:phot}.
The resulting noise became 0.56\,$\mu$Jy/beam.
The increase in noise should come from a contribution of the processing noise,
because we have not used real ISOCAM data in this case.
By subtracting the confusion noise above quadratically,
the processing noise was estimated to be 0.28\,$\mu$Jy/beam.

As the total noise of the real ISOCAM data is 1.1\,$\mu$Jy/beam,
the detector noise in the ISOCAM data should be 0.91\,$\mu$Jy/beam
by a quadratical subtraction.
It assures that our observations were dominated by the detector noise,
i.e.\ readout and photon noises.


Note: HTTP = http://www.iso.vilspa.esa.es

\end{document}

%% file: 1850_t3.tex
 0 & 13 12 15.8 & +42 43 60 & 2.5 & ?2.7 & $?19\;^{+11}_{- 9}$ & $0.77\;^{1.03}_{0.59}$ & 0.24 \\
 1 & 13 12 16.8 & +42 44 21 & 1.9 & ?1.7 & $?19\;^{+ 8}_{- 7}$ & $0.77\;^{1.01}_{0.61}$ & 0.56 \\
 2 & 13 12 17.4 & +42 44 53 & 2.3 & ?1.7 & $?14\;^{+ 7}_{- 6}$ & $0.77\;^{1.00}_{0.60}$ & 0.32 \\
 3 & 13 12 18.0 & +42 43 45 & 1.1 & ?1.4 & $?50\;^{+13}_{-12}$ & $0.90\;^{1.06}_{0.74}$ & 0.98 \\
 4 & 13 12 18.1 & +42 45 28 & 2.3 & ?2.2 & $?18\;^{+ 9}_{- 8}$ & $0.75\;^{1.01}_{0.59}$ & 0.36 \\
 5 & 13 12 18.1 & +42 46 15 & 2.6 & ?4.9 & $?35\;^{+19}_{-17}$ & $0.77\;^{1.03}_{0.60}$ & 0.24 \\
 6 & 13 12 18.4 & +42 43 20 & 1.1 & ?1.8 & $?66\;^{+17}_{-16}$ & $0.91\;^{1.06}_{0.74}$ & 0.98 \\
 7 & 13 12 18.7 & +42 42 48 & 2.3 & ?8.2 & $?66\;^{+34}_{-30}$ & $0.78\;^{1.02}_{0.61}$ & 0.33 \\
 8 & 13 12 19.5 & +42 44 60 & 1.9 & ?1.1 & $?12\;^{+ 6}_{- 5}$ & $0.78\;^{1.02}_{0.61}$ & 0.55 \\
 9 & 13 12 19.5 & +42 45 37 & 1.5 & ?1.5 & $?25\;^{+ 8}_{- 8}$ & $0.85\;^{1.04}_{0.67}$ & 0.86 \\
10 & 13 12 20.0 & +42 44 38 & 1.4 & ?1.1 & $?21\;^{+ 7}_{- 7}$ & $0.84\;^{1.06}_{0.65}$ & 0.89 \\
11 & 13 12 21.0 & +42 44 33 & 1.5 & ?1.1 & $?19\;^{+ 6}_{- 6}$ & $0.82\;^{1.01}_{0.65}$ & 0.85 \\
12 & 13 12 21.3 & +42 46 24 & 2.3 & ?7.9 & $?63\;^{+32}_{-28}$ & $0.77\;^{1.00}_{0.60}$ & 0.32 \\
13 & 13 12 21.4 & +42 44 24 & 1.4 & ?1.1 & $?24\;^{+ 7}_{- 7}$ & $0.87\;^{1.06}_{0.68}$ & 0.92 \\
14 & 13 12 21.6 & +42 44 05 & 1.5 & ?1.1 & $?19\;^{+ 7}_{- 6}$ & $0.82\;^{1.03}_{0.64}$ & 0.88 \\
15 & 13 12 21.6 & +42 45 19 & 1.0 & ?1.2 & $?60\;^{+13}_{-14}$ & $0.94\;^{1.06}_{0.78}$ & 1.00 \\
16 & 13 12 21.8 & +42 43 48 & 1.5 & ?1.1 & $?19\;^{+ 6}_{- 6}$ & $0.84\;^{1.03}_{0.67}$ & 0.86 \\
17 & 13 12 22.6 & +42 44 51 & 1.5 & ?1.1 & $?17\;^{+ 6}_{- 6}$ & $0.84\;^{1.07}_{0.64}$ & 0.80 \\
18 & 13 12 22.8 & +42 43 15 & 2.7 & ?1.2 & $??8\;^{+ 5}_{- 4}$ & $0.75\;^{0.98}_{0.60}$ & 0.17 \\
19 & 13 12 23.1 & +42 46 07 & 2.2 & ?1.9 & $?17\;^{+ 9}_{- 7}$ & $0.79\;^{1.07}_{0.59}$ & 0.42 \\
20 & 13 12 23.4 & +42 45 17 & 1.5 & ?1.1 & $?20\;^{+ 7}_{- 6}$ & $0.82\;^{1.02}_{0.65}$ & 0.87 \\
21 & 13 12 23.4 & +42 44 31 & 2.3 & ?1.1 & $??9\;^{+ 4}_{- 4}$ & $0.78\;^{1.00}_{0.61}$ & 0.34 \\
22 & 13 12 24.0 & +42 45 43 & 0.9 & ?1.4 & $135\;^{+24}_{-22}$ & $0.97\;^{1.04}_{0.90}$ & 1.00 \\
23 & 13 12 24.3 & +42 43 59 & 2.3 & ?1.1 & $??9\;^{+ 4}_{- 4}$ & $0.76\;^{0.98}_{0.61}$ & 0.33 \\
24 & 13 12 24.5 & +42 43 23 & 2.0 & ?1.1 & $?12\;^{+ 5}_{- 5}$ & $0.77\;^{1.01}_{0.59}$ & 0.50 \\
25 & 13 12 24.7 & +42 44 15 & 1.1 & ?1.1 & $?39\;^{+10}_{- 9}$ & $0.90\;^{1.05}_{0.75}$ & 0.98 \\
26 & 13 12 24.8 & +42 45 10 & 2.1 & ?1.1 & $?10\;^{+ 5}_{- 4}$ & $0.76\;^{1.02}_{0.59}$ & 0.44 \\
27 & 13 12 25.2 & +42 46 00 & 1.2 & ?1.4 & $?45\;^{+12}_{-12}$ & $0.91\;^{1.07}_{0.74}$ & 0.97 \\
28 & 13 12 25.3 & +42 43 46 & 1.3 & ?1.1 & $?27\;^{+ 8}_{- 8}$ & $0.85\;^{1.04}_{0.66}$ & 0.95 \\
29 & 13 12 25.7 & +42 43 19 & 2.4 & ?1.1 & $??8\;^{+ 5}_{- 4}$ & $0.77\;^{1.04}_{0.60}$ & 0.27 \\

%% file: 1850_t3+.tex
30 & 13 12 26.2 & +42 42 27 & 1.6 & ?4.9 & $?70\;^{+28}_{-25}$ & $0.82\;^{1.06}_{0.63}$ & 0.75 \\
31 & 13 12 26.3 & +42 45 26 & 3.2 & ?1.1 & $??6\;^{+ 3}_{- 3}$ & $0.72\;^{0.86}_{0.59}$ & 0.08 \\
32 & 13 12 26.3 & +42 44 06 & 2.1 & ?1.0 & $??9\;^{+ 5}_{- 4}$ & $0.78\;^{1.06}_{0.59}$ & 0.43 \\
33 & 13 12 27.4 & +42 44 49 & 1.5 & ?1.1 & $?17\;^{+ 6}_{- 6}$ & $0.84\;^{1.07}_{0.65}$ & 0.82 \\
34 & 13 12 27.7 & +42 46 32 & 2.4 & ?5.9 & $?45\;^{+23}_{-21}$ & $0.76\;^{0.98}_{0.59}$ & 0.30 \\
35 & 13 12 27.7 & +42 45 35 & 1.7 & ?1.1 & $?15\;^{+ 6}_{- 5}$ & $0.80\;^{1.03}_{0.62}$ & 0.71 \\
36 & 13 12 27.7 & +42 45 09 & 2.4 & ?1.1 & $??8\;^{+ 4}_{- 4}$ & $0.76\;^{1.02}_{0.59}$ & 0.27 \\
37 & 13 12 27.9 & +42 42 20 & 2.4 & 11.8 & $?87\;^{+45}_{-40}$ & $0.74\;^{0.97}_{0.58}$ & 0.27 \\
38 & 13 12 28.3 & +42 44 30 & 2.1 & ?1.0 & $?10\;^{+ 5}_{- 4}$ & $0.75\;^{1.01}_{0.59}$ & 0.45 \\
39 & 13 12 28.5 & +42 46 02 & 1.5 & ?1.7 & $?27\;^{+10}_{- 9}$ & $0.86\;^{1.08}_{0.68}$ & 0.83 \\
40 & 13 12 28.5 & +42 44 52 & 1.7 & ?1.1 & $?14\;^{+ 6}_{- 5}$ & $0.80\;^{1.02}_{0.62}$ & 0.70 \\
41 & 13 12 28.6 & +42 43 60 & 1.1 & ?1.1 & $?46\;^{+10}_{-10}$ & $0.93\;^{1.05}_{0.78}$ & 0.99 \\
42 & 13 12 28.9 & +42 46 19 & 2.3 & ?3.8 & $?32\;^{+17}_{-14}$ & $0.78\;^{1.04}_{0.60}$ & 0.37 \\
43 & 13 12 29.0 & +42 43 07 & 2.1 & ?1.2 & $?12\;^{+ 6}_{- 5}$ & $0.75\;^{1.01}_{0.59}$ & 0.46 \\
44 & 13 12 29.2 & +42 45 59 & 1.3 & ?1.6 & $?37\;^{+12}_{-11}$ & $0.84\;^{1.03}_{0.65}$ & 0.95 \\
45 & 13 12 29.4 & +42 44 40 & 3.0 & ?1.1 & $??7\;^{+ 4}_{- 3}$ & $0.74\;^{0.92}_{0.61}$ & 0.12 \\
46 & 13 12 29.6 & +42 43 01 & 2.3 & ?1.3 & $?11\;^{+ 6}_{- 5}$ & $0.75\;^{1.00}_{0.59}$ & 0.36 \\
47 & 13 12 29.9 & +42 44 08 & 1.5 & ?1.0 & $?16\;^{+ 6}_{- 5}$ & $0.84\;^{1.07}_{0.64}$ & 0.80 \\
48 & 13 12 29.9 & +42 45 14 & 2.3 & ?1.1 & $??9\;^{+ 5}_{- 4}$ & $0.77\;^{1.02}_{0.61}$ & 0.37 \\
49 & 13 12 30.0 & +42 44 20 & 1.6 & ?1.0 & $?15\;^{+ 6}_{- 5}$ & $0.82\;^{1.05}_{0.63}$ & 0.75 \\
50 & 13 12 30.1 & +42 44 59 & 1.5 & ?1.1 & $?19\;^{+ 6}_{- 6}$ & $0.85\;^{1.04}_{0.67}$ & 0.86 \\
51 & 13 12 31.1 & +42 43 33 & 1.4 & ?1.1 & $?22\;^{+ 8}_{- 7}$ & $0.84\;^{1.06}_{0.64}$ & 0.91 \\
52 & 13 12 31.5 & +42 42 44 & 2.3 & ?3.0 & $?24\;^{+12}_{-10}$ & $0.75\;^{0.99}_{0.59}$ & 0.36 \\
53 & 13 12 31.5 & +42 42 36 & 2.9 & ?4.3 & $?27\;^{+15}_{-14}$ & $0.74\;^{0.94}_{0.60}$ & 0.11 \\
54 & 13 12 31.6 & +42 45 51 & 0.9 & ?2.1 & $167\;^{+32}_{-35}$ & $0.95\;^{1.05}_{0.81}$ & 1.00 \\
55 & 13 12 31.8 & +42 43 48 & 1.9 & ?1.0 & $?12\;^{+ 5}_{- 4}$ & $0.75\;^{0.98}_{0.59}$ & 0.58 \\
56 & 13 12 32.0 & +42 45 05 & 2.4 & ?1.1 & $??8\;^{+ 4}_{- 4}$ & $0.74\;^{0.97}_{0.58}$ & 0.27 \\
57 & 13 12 32.0 & +42 44 32 & 1.5 & ?1.0 & $?18\;^{+ 6}_{- 6}$ & $0.83\;^{1.02}_{0.66}$ & 0.88 \\
58 & 13 12 32.9 & +42 43 11 & 2.6 & ?2.0 & $?15\;^{+ 8}_{- 7}$ & $0.76\;^{1.02}_{0.60}$ & 0.24 \\
59 & 13 12 34.0 & +42 44 42 & 1.6 & ?1.3 & $?18\;^{+ 7}_{- 6}$ & $0.80\;^{1.03}_{0.62}$ & 0.74 \\
60 & 13 12 34.2 & +42 46 02 & 3.3 & ?4.4 & $?26\;^{+16}_{-14}$ & $0.72\;^{0.97}_{0.58}$ & 0.07 \\
61 & 13 12 34.5 & +42 43 41 & 1.3 & ?2.3 & $?55\;^{+16}_{-16}$ & $0.85\;^{1.03}_{0.66}$ & 0.95 \\
62 & 13 12 34.7 & +42 43 08 & 1.8 & ?7.1 & $?88\;^{+36}_{-31}$ & $0.80\;^{1.03}_{0.63}$ & 0.66 \\
63 & 13 12 35.0 & +42 45 05 & 2.1 & ?1.9 & $?18\;^{+ 9}_{- 7}$ & $0.76\;^{1.01}_{0.59}$ & 0.48 \\
64 & 13 12 35.8 & +42 43 60 & 1.3 & ?4.7 & $107\;^{+34}_{-33}$ & $0.85\;^{1.06}_{0.65}$ & 0.94 \\